\documentclass[a4paper]{article}

\usepackage{titling}
\usepackage[utf8]{inputenc}
\usepackage[T1]{fontenc}
\usepackage{textgreek}

\usepackage[a4paper]{geometry}
\usepackage{amsmath,stackengine,mathtools,amssymb}
\usepackage{amsthm} 
\usepackage{comment}
\usepackage{lipsum}
\usepackage[usenames,dvipsnames]{xcolor}
\usepackage{minitoc}
\usepackage{tikz}
\usepackage{filecontents}
\usepackage[numbers,sort&compress]{natbib}
\usepackage[colorlinks=true]{hyperref} 
\hypersetup{urlcolor=blue, citecolor=red} 
\usepackage{tikz} 
\usetikzlibrary{matrix} 
\usepackage{graphicx}
\usepackage[english]{babel}
\usepackage{hyperref}
\usepackage{amsfonts}
\usepackage{empheq} 
\usepackage{tabularx}
\usepackage{multirow}
\usepackage{appendix}
\usepackage{color} 
\usepackage{subcaption}
\usepackage{lineno} 
\usepackage{float} 
\usepackage{caption} 
\usepackage{xcolor}   
\usepackage{environ}  
\usepackage{enumitem}       
\usepackage{booktabs}       
\usepackage{diagbox} 
\usepackage{array} 
\usepackage{natbib} 
\usepackage{soul,xcolor} 
\usepackage{cancel}
\usepackage{siunitx} 
\newcommand\Tstrut{\rule{0pt}{2.6ex}} 
\newcommand\Bstrut{\rule[-0.9ex]{0pt}{0pt}} 

\newtheorem{assump}{Assumption}[section]

\newtheorem{remark}{Remark}[section]

\newcommand{\tr}[1]{{#1}}
\definecolor{newgreen}{rgb}{0.0, 0.5, 0.0}

\newcommand{\height}{h}
\newcommand{\coh}{\mathbf{{c}}}
\newcommand{\commandtitle}{
	Deciphering circulating tumor cells binding in a microfluidic system thanks to a parameterized mathematical model
}
\usepackage{bigfoot} 
\usepackage[numbered,framed]{matlab-prettifier} 

\usepackage{natbib}

\usepackage{slashbox}

\usepackage{filecontents} 

\lstMakeShortInline"

\makeatletter
\let\@fnsymbol\@arabic
\makeatother

\title{\commandtitle}

\author{Giorgia Ciavolella\thanks{Institut Denis Poisson, Université d’Orléans, CNRS, Université de Tours, 45067 Orléans, France}	\and
	Julien Granet\protect\thanks{Inria, Univ. Bordeaux, CNRS, Bordeaux INP, IMB, UMR 5251, F-33400 Talence, France}
	\and
	Jacky G. Goetz \thanks{INSERM UMR\_S 1109, Univ. Strasbourg, FMTS, \'Equipe labellis\'ee Ligue Contre le Cancer, F-67000 Strasbourg, France.}
	\and
	Na{\"e}l Osmani\protect\footnotemark[3]
	\and
	Christ\`ele Etchegaray\protect\footnotemark[2] 
	\and
	Annabelle Collin\protect\thanks{Laboratoire de Math\'ematiques Jean Leray, Nantes Universit\'e, F-44100 Nantes, France} }

\date{\today}
\begin{document} 

\maketitle

\begin{abstract} 
The spread of metastases is a crucial process in which some questions remain unanswered. In this work, we focus on tumor cells circulating in the bloodstream, the so-called Circulating Tumor Cells (CTCs). Our aim is to characterize their trajectories under the influence of hemodynamic and adhesion forces.
We focus on already available \emph{in vitro} measurements performed with a microfluidic device corresponding to the trajectories of CTCs --~without or with different protein depletions~-- interacting with an endothelial layer.  A key difficulty is the weak knowledge of the fluid velocity that has to be reconstructed. Our strategy combines a differential equation model --~ a Poiseuille model for the fluid velocity and an ODE system for the cell adhesion model~-- and a robust and well-designed calibration procedure.  The parameterized model quantifies the strong influence of fluid velocity on adhesion and confirms the expected role of several proteins in the deceleration of CTCs.
Finally, it enables the generation of synthetic cells, even for unobserved experimental conditions, opening the way to a digital twin for flowing cells with adhesion.
\end{abstract}

\begin{flushleft}
	\noindent{\makebox[1in]\hrulefill}
\end{flushleft}
\\[-2.em]
\begin{flushright}
	\noindent{\makebox[1in]\hrulefill}
\end{flushright}

\section{Introduction}
One of the most important and deadly features of solid tumors is the increased ability of cancer cells to migrate and invade other organs, which is called metastatic spread. 
In the last $70$ years the number of cancer deaths registered with metastasis has tripled~\cite{Dillekas}. Although the incidence of individual types of cancer varies greatly, metastasis remains the main cause of cancer-related deaths~\cite{steeg2016targeting}. 

The blood and lymphatic circulations are used as a means of transport to reach distant organs. Tumor cells that have previously detached from a primary tumor can invade the surrounding extracellular matrix. Successful intravasation into the vessels means that cancer cells can now leave the original site. Inside the blood vessels, hostile conditions prevail. Circulating Tumor Cells (CTCs) are subjected to physical stresses that include hydrodynamic flow and loss of attachment to a substrate, as well as other obstacles involving the human immune system (and platelets)~\cite{wirtz, follain2}. These factors lead to a significant decrease in the number of CTCs and also to their eventual clustering. The remaining single cells or small cell clusters eventually extravasate, reaching a secondary site where they either stay dormant or form a new tumor~\cite{obenauf}.

CTCs receive much research interest due to their therapeutic potential in liquid biopsy~\cite{lim_liquid_2019, peralta}. Indeed, they could allow to monitor tumor heterogeneity or response to a treatment, but also to detect the minimal residual disease, and serve as a prognosis biomarker or as a target for personalized therapies~\cite{joosse15, liu_circulating_2021, ring_biology_2023}. However, the detection, identification and characterization of CTCs present important challenges due to their heterogeneity and low abundance~\cite{huang_sensitive_2019}. From the biological standpoint, understanding the key steps involved in CTCs arrest on the endothelial wall is crucial to explain secondary tumour locations. Indeed, the possibility of extravasating is permitted by CTCs arrest and firm adhesion to the vascular endothelium, phenomena that need further insights~\cite{rejniak_circulating_2016,follain,osmani,paul_tissue_2019,martinez-pena_dissecting_2021}.

Studies previously pursued by biologists Follain et al.~\cite{follain} and Osmani et al.~\cite{osmani} have deepened into the mechanical cues that promote CTCs successful arrest and extravasation {using statistical approaches}. In~\cite{follain}, they have shown that an optimal flow is required for CTCs to arrest on the endothelium of the vascular wall. Furthermore, in~\cite{osmani}, they have identified the adhesion receptors at play. Early adhesion is mediated by the glycoprotein CD44, involved in a weak form of bonds, while integrin ITGB1 favours stabilization of the adhesions. {These findings originate from both} \textit{in vitro} and \textit{in vivo} experiments { on the metastatic D2A1 cell line, and were validated in a fully syngeneic experimental mouse metastasis model}.  {The} \textit{in vitro} experiments consist in using a microfluidic channel with controlled fluid velocity (simulating a blood vessel) into which tumor cells are injected.  {They allowed to establish that both CD44 and ITGB1 favour cell arrest, while only ITGB1 is necessary for stable cell adhesion.  The resulting data were validated \textit{in vivo} using intravital imaging in} zebrafish embryos where CTCs {were} pumped by the heart along the vascular architecture.  {In that setting, it was further proved that CD44 mediates early cell arrest (5 minutes after cell injection), while ITGB1 is mainly required for stable cell adhesion (3 hours after injection)}. {These two molecules were identified \textit{in vitro} among several others such as ITGA3 and ITGA4 that were not found significant, and ITGA5 that was identified as the partner subunit of ITGB1. Furthermore, ITGB3 was found to behave as CD44 in mediating cell arrest.} {In summary, these recent works point out the interplay of the flow and the CD44/ITGB3 and ITGB1{/ITGA5} molecules on CTCs transient or stable adhesion to the endothelial wall. However, they do not inform on how their whole dynamics may be affected even before arrest.} In the present work,  {we aim to exploit the \emph{in vitro} data from~\cite{osmani} to derive cell trajectories and to interpret them in the light of \textbf{deterministic mathematical modeling}.  This will allow to decipher the respective \textbf{roles of the shear flow} and of the identified molecules of interest in the early \textbf{interaction between CTCs and the endothelial wall}.}

Various theoretical models of cell adhesion have been developed over time. First studies have focused on the binding dynamics of {either single bonds or clusters sharing a constant or varying load~\cite{bell_78,bell1984cell,erdmann,erdmann_adhesion_2004,erdmann_impact_2007}}. In the case of inflow cell dynamics, several biological questions can be addressed, such as the emergence of several cell displacement regimes (freely-flowing, rolling, slipping, stationary arrest) with possible bistability or shear-threshold effect between them. A related issue concerns the bonds response to hydrodynamic forces, with \textit{catch bonds} whose lifetime increases with load, \textit{slip bonds} for which it decreases exponentially with load (so-called Bell's law), or a combination of both depending on the shear rate.

First computational approaches allowed to describe a hard sphere submitted to hydrodynamic forces and stochastic binding interaction with the wall~\cite{hammer_dynamical_1987,dembo,hammer_simulation_1992,chang_state_2000,caputo_adhesive_2007,beste_pnas_2008,efremov_bistability_2011}.  In~\cite{li,ijms21020584}, the adhesion of a rolling sphere is described following the membrane approaching the wall at the front, and detaching at the rear, for catch-slip bonds. 
In~\cite{korn_dynamic_2008}, both translational and rotational motions of a spherical cell are affected by elastic bonds. This allows to explain the interplay between rolling and slipping, and provides a numerical state diagram of leukocyte motion. Theoretical models can also take ligands positions into account, thus enabling bonds tilting and subsequent cell sliding~\cite{reboux,PreziosiVitale11,milisic_asymptotic_2011,milisic_structured_2015,grec}. {In the spirit of \cite{grec}}
but in the absence of space structure, stochastic and deterministic models are developed for a particle cell in~\cite{etchegaray}. 
Although minimal in the hydrodynamic description, these models have less parameters and are therefore more suited to calibration with experimental data.

Theoretical frameworks have been confronted to microfluidics experiments on CTCs in~\cite{cheung_detachment_2009,Cheung10,cheung_adhesion_2011}.
In particular, in~\cite{cheung_adhesion_2011}, the authors perform microfluidics experiments to study the effect of the shear rate on the dynamics of breast cancer cells interacting with an EpCAM-coated wall. Three regimes were observed (freely-flowing, firmly adhering, and rolling/slipping). Experimental data consisted in trajectories and in stopping times and lengths that were used to empirically calibrate a model based on~\cite{korn_dynamic_2008}. More precisely, the cell-wall gap, the typical adhesion force and the spring constant were sequentially identified by numerical investigations. Then, the cell velocity during capture was well fitted by a decreasing exponential function, yielding a typical decreasing time characteristic of the cell-wall interaction.

In this work, we aim to capture the role of hydrodynamic forces and of {different} adhesion proteins {during} the first phase of CTCs interaction with the endothelial wall {using a mathematical model.} 
{First, we exploit the movies from the \emph{in vitro} experiments from~\cite{osmani} and we derive cell trajectories by tracking. Next, we perform a statistical analysis of the resulting mean velocities to highlight significantly different mean dynamics. Then, we build a mathematical model to explain these differences, relying on the full reconstruction of the fluid velocity}. We {consider} a Poiseuille model for the fluid velocity, and weakly couple it to a modification of the adhesion model proposed in~\cite{grec,etchegaray}. The cell velocity depends on both the fluid velocity and the bonds density, while the binding dynamics takes into account bonds formation, adhesion growth, and unbinding. A well-designed and robust calibration based on nonlinear mixed-effects models~\cite{lavielle2014mixed} is used to fit the model to the \textit{in vitro} experiments carried out by Osmani and collaborators, see~\cite{osmani,osmani2}.

The work is arranged as follows. Section~\ref{sec:data} contains the main information about the biological data: see Subsection~\ref{subsec:protocol} for protocol details and Subsection~\ref{subsec:data_availability} for data presentation. Trajectories and velocities {of $278$} cells with or without proteins depletions were extracted from these data. Section~\ref{sec:methods} is devoted to methods. After a brief statistical analysis of the data in Subsection~\ref{subsec:statisticalStudy}, which shows the statistically significant slowing behavior of CTC velocities over time in most cases, Subsections~\ref{subsec:fluid_velocity} and~\ref{subsec:cells_velocity} present the mathematical modeling. In Subsection~\ref{subsec:param_estimation}, the parameter estimation strategy is presented.  {The strategy to obtain a digital twin after calibration is presented in Subsection~\ref{subsec:digital_twin}.} The results showing the good agreement of the model with the data {and the obtained synthetic cells} are presented in Section~\ref{sec:results}. A discussion is presented in Section~\ref{sec:discussion}, and finally, conclusions are given in Section~\ref{sec:conclusions}. 

\section{Data}\label{sec:data}

In this section, we present the experimental data, beginning with their acquisition and ending with their extraction. First, in Subsection~\ref{subsec:protocol} we present the experimental protocol. Then, in Subsection~\ref{subsec:data_availability}, we show what kind of data are obtained, and briefly present the tracking techniques used to extract the trajectory and velocity of 278 cells. We then present the resulting cell velocities.

\subsection{Protocol} \label{subsec:protocol}
{This study relies on the experimental \textit{in vitro} data obtained in~\cite{osmani},  to which we refer to for further details}. Human Umbilical Vein Endothelial  cells (\textbf{HUVEC}, Promocell) were seeded at 30 000 cells per channel in a rectangular microfluidic channel (IBIDI) of length $L = \SI{1.7d-2}{\meter}$, of width $l = \SI{3.8d-3}{\meter}$ and height $h = \SI{4.0d-4}{\meter}$. Medium was changed twice a day until they reach maximal confluency (3 to 4 days). \textbf{DA21} mouse breast carcinoma cells were with siRNA using Lipofectamine RNAiMAX (Thermo Fisher) following the manufacturer’s instructions. Experiments were performed between 72 hours and 96 hours post-transfection. 3 days after siRNA transfection, \textbf{DA21} cells were resuspended at a concentration of $10^{6}$ cells/ml in a Hépès-buffered cell culture medium and perfused into the channel using  using a REGLO Digital MS-2/12 peristaltic pump (Ismatec), Tygon LMT-55 3-stop tubing (IDEX), 0.5 and 1.6 mm silicon tubing and elbow Luer connectors (IBIDI). 

\begin{figure}
	\begin{center}	
		\includegraphics[width=\textwidth]{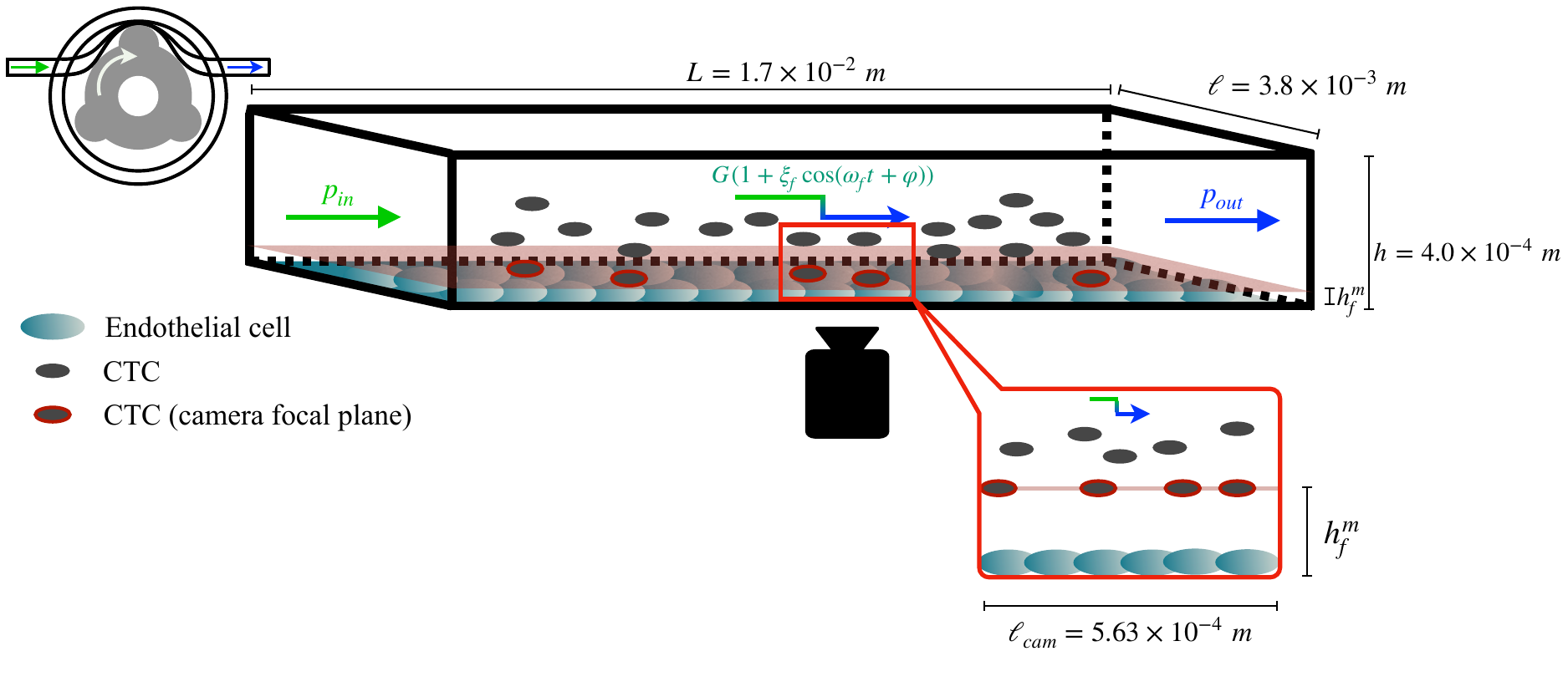}
		\caption{Experimental setup. A rectangular microfluidic channel (black box) contains an endothelial layer (turquoise cells) and a fluid containing \textbf{DA21} tumor cells (grey cells). The fluid is pumped by a peristaltic pump (left in the figure) with a pressure gradient $G(1+\xi_f \cos(\omega_f t +\varphi))$ at the entrance. A camera is placed to record the motion of cells located in a focal plane at a distance $\height_f^m$ from the endothelial layer (grey cells surrounded by a red contour), see~\cite{osmani2}. }
		\label{fig:setup}
	\end{center}
\end{figure}

In the setup of the pump, the mean value of the entry pressure gradient is fixed and denoted by $G$. The fluid velocity generated --~which contains oscillations due to the pump~-- depends on the position in the channel and is not measured. A cMOS camera (IDS) is placed to record the motion of cells located in a focal plane at a distance $\height_f^m$ from the endothelial layer. The experimental data consist of timelapse movie acquired at a rate of 24 frames per seconds for 2 minutes, on a rectangle of width $\ell_{cam} = \SI{5.63d-4}{\meter}$ and of height $h_{cam} = \SI{2.99d-4}{\meter}$. 
The setup is shown in Figure~\ref{fig:setup}.

\subsection{Data availability} \label{subsec:data_availability}

\begin{figure}	
	\begin{center}	
		\includegraphics[width= .8\textwidth]{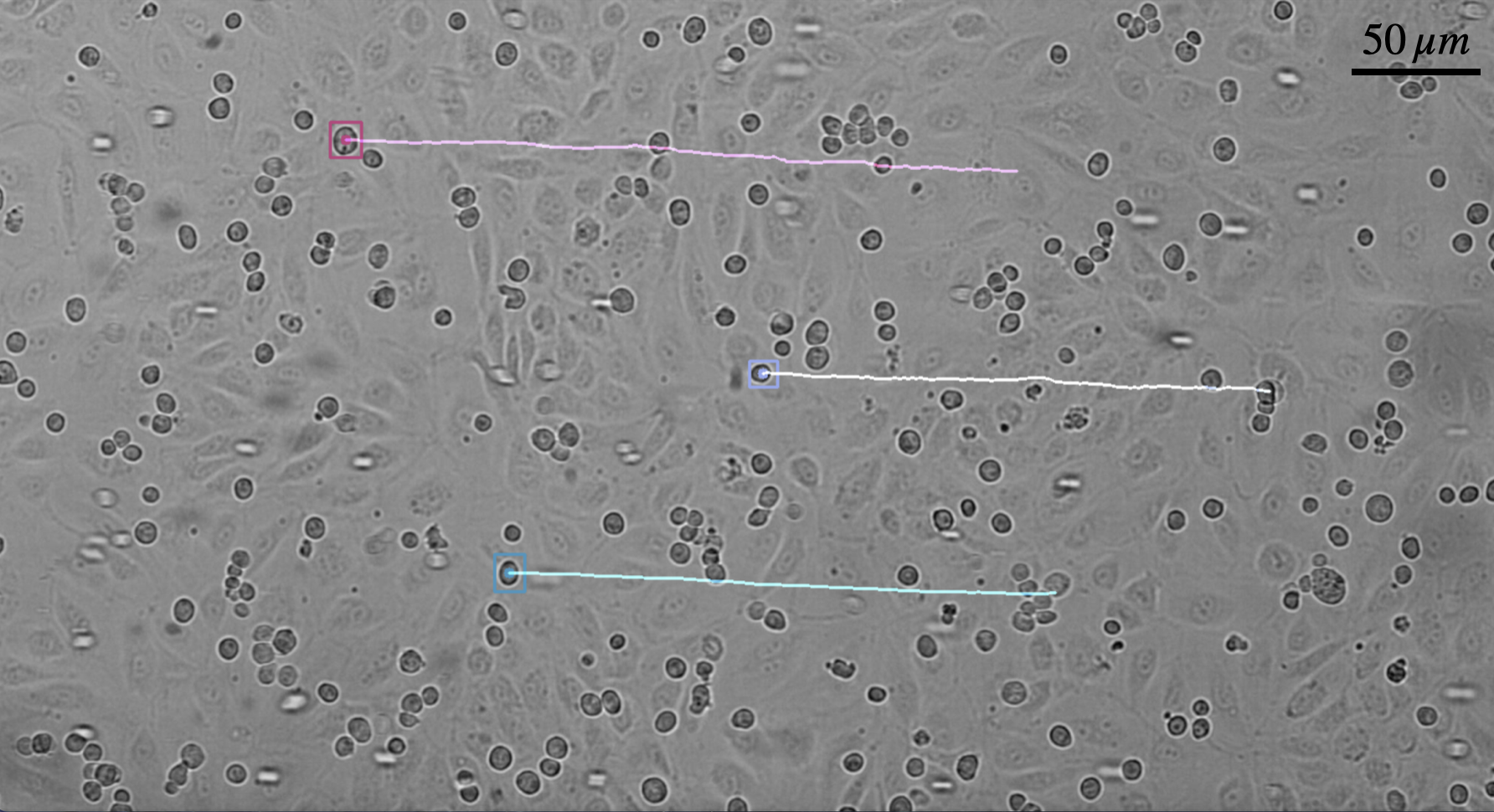}
		\caption{Example of the use of the CSRT tracker on three different cells of the video corresponding to the control case (siCTL) {of Experiment 1,} at a fluid velocity corresponding to the smallest pressure gradient $G^{(1)} = \SI{50.33}{\pascal \per \meter}$.} 
		\label{fig:trajectory}
	\end{center}
\end{figure}

An example of a video image is shown in Figure~\ref{fig:trajectory}. The cells forming the endothelial layer are seen in the background, whereas moving CTCs that are not in the focal plane of the camera are seen in the foreground. These CTCs have a well-defined shape, so that their trajectories can easily be followed while they appear in the video. Most cells are smoothly transported through the fluid. Sometimes, a cell stops on the endothelial layer. This arrest can be stable, meaning that the cell remains attached to the endothelial layer, or unstable, \textit{i.e.} the cell can detach due to a collision with other cells or under the effect of the hydrodynamic forces. However, in this work we focus only on non-arrested cells.

{Three experiments are considered, each of them marked by a control case representing the unmodified biological conditions (D2A1 cells treated with a CTL siRNA):
\begin{itemize} \setlength\itemsep{-0.5em}
	\item[(1)] Experiment 1: Case ITGB1, CD44
	\begin{itemize} \setlength\itemsep{0em}
		\item[$\bullet$] siCTL (ITGB1, CD44): control group;
		\item[$\bullet$] siITGB1: depletion of ITGB1 (D2A1 cells treated with a siRNA targeting ITGB1);
		\item[$\bullet$] siCD44: depletion of CD44 (D2A1 cells treated with a siRNA targeting CD44);
	\end{itemize}
	\item[(2)] Experiment 2: Case ITGA5
	\begin{itemize} \setlength\itemsep{0em}
	\item[$\bullet$] siCTL (ITGA5): control group;
	\item[$\bullet$] siITGA5: depletion of ITGA5 (D2A1 cells treated with a siRNA targeting ITGA5);
	\end{itemize}
	\item[(3)] Experiment 3: Case ITGB3
	\begin{itemize} \setlength\itemsep{0em}
	\item[$\bullet$] siCTL (ITGB3): control group;
	\item[$\bullet$] siITGB3: depletion of ITGB3 (D2A1 cells treated with a siRNA targeting ITGB3)
	\end{itemize}
\end{itemize} 
}

The different experimental subgroups are summarized in Table~\ref{table:cells_number}. Experiments have been realized keeping the fluid at a controlled pressure gradient by the peristaltic pump. Three possible values of pressure gradient cohorts have been considered: $G^{(1)} = \SI{50.33}{\pascal \per \meter}$, $G^{(2)} = \SI{100.66}{\pascal \per \meter}$ and $G^{(3)} = \SI{201.32}{\pascal \per \meter}$.  For each of these cohorts, control and depletion of adhesion proteins gives rise to seven sub-cohorts, see Western-blot results in Supplementary materials in~\cite{osmani}.

We collect cell trajectories from the ${17}$ different videos {(Experiment~$1$: $3$ different pressure gradients cohorts $\times$ $3$ different proteins expressions sub-cohorts; Experiment $2$: $2$ cohorts $\times$ $2$ sub-cohorts; ; Experiment $3$: $2$ cohorts $\times$ $2$ sub-cohorts).} {A preliminary analysis of data in Experiment~$1$ and the discovery of not reliable and significative results for the higher pressure gradient cohort are the reason why in the other two experiments we have not considered this value. } For the two first values of the pressure gradient, we use a semi-automatic tracker called \textit{Channel and Spatial Reliability Tracker} (CSRT) that consists in first manually designing a box surrounding the cell of interest and second automatically recording the box evolution at each frame of the video~\cite{lukezic}, see Figure~\ref{fig:trajectory}.
In the videos realised with the highest pressure gradient, the mean fluid velocity is too high for the tracker to automatically track the CTCs. For this reason, they were tracked manually.  
By deriving the trajectories using a first-order scheme, we can directly determine the cell velocities. Note that the tracking procedure captures only the translational motion. Therefore, the data used in this work do not allow discussing possible CTCs rolling.

\begin{table}
	\begin{subtable}{\textwidth}
		\centering
		\caption{\textbf{Experiment $\mathbf{1}$}}
		\begin{tabular}{|c|c|c|c|c|} 
			\hline
			\backslashbox{$G$ }{Protein modification} & {\textbf{siCTL}} & {\textbf{siITGB1}} & {\textbf{siCD44}} & Total  \\ 
			\hline
			{\bfseries $\SI{50.33}{\pascal \per \meter}$} & $15$ $(15)$  & $15$ $(16)$  & $12 $ $(14)$ & $\mathbf{42}$ $(45)$ \\
			{\bfseries $\SI{100.66}{\pascal \per \meter}$} & $14$  & $20$  & $9 $ $(11)$ & $\mathbf{43}$ $(45)$\\
			{\bfseries $\SI{201.32}{\pascal \per \meter}$} & $24 $ $(29)$  & $14$  &  $14$ $ (16)$ & $\mathbf{52}$ $(59)$\\
			\hline
			Total & $\mathbf{53}$ $(58)$  & $\mathbf{49}$ $(50)$   & $\mathbf{35}$ $(41)$ &  $\mathbf{137}$ $(149)$ \\
			\hline
		\end{tabular}
	\end{subtable}
	\begin{subtable}{\textwidth}
		\centering
		\caption{\textbf{Experiment $\mathbf{2}$}}
		\begin{tabular}{|c|c|c|c|} 
			\hline
			\backslashbox{$G$ }{Protein modification} & {\textbf{siCTL}} & {\textbf{siITGA5}}& Total  \\ 
			\hline
			{\bfseries $\SI{50.33}{\pascal \per \meter}$} & $11$ $(17)$  & $12$ $(14)$  & $\mathbf{23}$ $(31)$ \\
			{\bfseries $\SI{100.66}{\pascal \per \meter}$} & $7$  $(11)$& $18$  $(19)$& $\mathbf{25}$ $(30)$\\
			\hline
			Total & $\mathbf{18}$ $(28)$  & $\mathbf{30}$ $(33)$ & $\mathbf{48}$ $(61)$ \\
			\hline
		\end{tabular}
	\end{subtable}
	\begin{subtable}{\textwidth}
		\centering
		\caption{\textbf{Experiment $\mathbf{3}$}}
		\begin{tabular}{|c|c|c|c|} 
			\hline
			\backslashbox{$G$ }{Protein modification} & {\textbf{siCTL}} & {\textbf{siITGB3}}& Total  \\ 
			\hline
			{\bfseries $\SI{50.33}{\pascal \per \meter}$} & $11$ $(14)$  & $13$ $(19)$  & $\mathbf{24}$ $(33)$ \\
			{\bfseries $\SI{100.66}{\pascal \per \meter}$} & $6$ $(12)$ & $21$  $(23)$& $\mathbf{27}$ $(35)$\\
			\hline
			Total & $\mathbf{17}$ $(26)$  & $\mathbf{34}$ $(42)$ & $\mathbf{51}$ $(68)$ \\
			\hline
		\end{tabular}
	\end{subtable}
	\caption{Total number of cells considered in each {experiment}, cohort and sub-cohort after removing the outliers. The values in parenthesis correspond to the number of tracked cells.}
	\label{table:cells_number}
\end{table}

\begin{figure}
	\begin{center}	
		\includegraphics[width=\textwidth,trim = {0cm 0cm 0cm 0cm},clip]{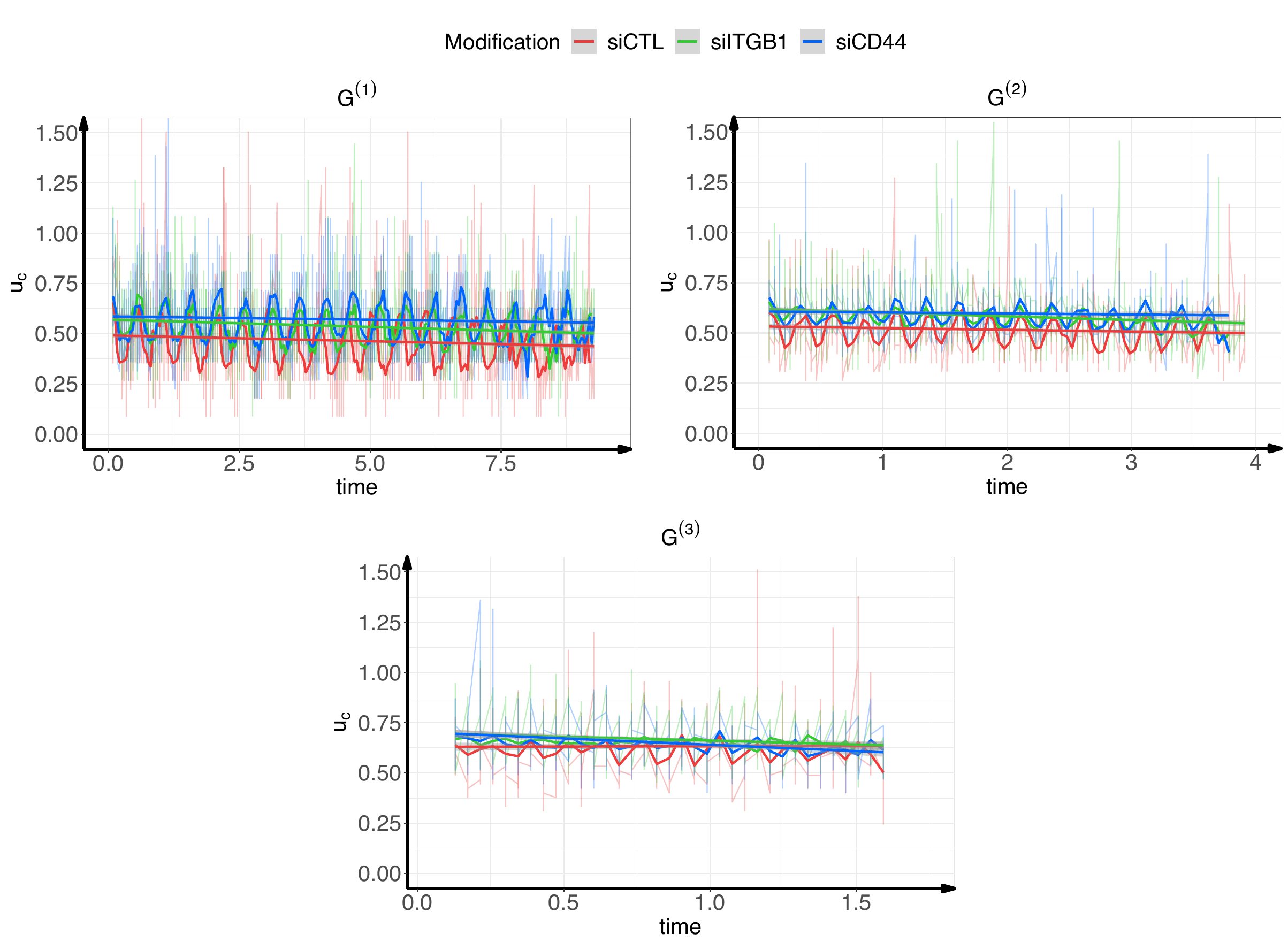}
		\caption{All extracted cell velocities over time {in Experiment 1} ($137$ cells, without outliers). Time is given in seconds, and velocities in micrometers per seconds. Cell velocities are given in transparent color and weighted mean across cells is shown in normal color. The choice of weighted means is motivated by a better visualisation of the results, since the noise is reduced.
			The three figures correspond to the three different pressure gradient cohorts: top left for $G^{(1)}$, top right for $G^{(2)}$ and bottom for $G^{(3)}$.
			The red curves stand for the siCTL cases, green for siITGB1 cases, and blue for siCD44 cases. For each figure, the straight lines correspond to the linear regressions of the weighted means of cell velocities. The initial time of each cell velocity is adjusted in order to synchronize the oscillations. The data are normalized respectively by $\SIlist{100;200;400}{\micro\meter\per\second}$. }
		\label{fig:vel_mean}
	\end{center}
\end{figure}

Table~\ref{table:cells_number} summarizes the number of cells considered in each of the {$17$} videos. For each velocity cohort, at least {$20$} cells were considered. Some cells were significantly different from the others since they had very high initial velocities. We assumed that they either collided with another cell before entering the video time-lapse, or that they were located in a different focal plane, and we excluded them. These outliers make up only a small portion of the data, since they counted for at most {$9$ for each velocity cohort ($42$ for $278$ cells $\sim 15 \%$)}, see Table~\ref{table:cells_number} for details. 

The extracted velocities over time cells are given in Figure~\ref{fig:vel_mean} {for Experiment~$1$, as an example, thus for $137$ cells out of $236$}. {Similar pictures can be obtained for the other two experiments.} Individual cell velocities are in transparent color and weighted mean across cells is shown in normal color. The choice of weighted means is motivated by a better visualisation of the results, since the noise is reduced. Indeed, the raw data are preprocessed as follows. First, the velocities are normalized with respect to the fluid pressure cohort, to remove the linear dependence on the fluid velocity. Each velocity in transparent color is corrected for phase shift to synchronize the oscillations. To do so, we use the estimated cell velocities aligning their maximum points over two periods of oscillation. Finally, spurious data are filtered, to deal with the additional noise brought by the first-order derivation of velocities from positions. Indeed, some velocity values can be artificially large or low and perturb the analysis. To deal with this difficulty, we compute weighted means assigning a null weight on values above $\SI{0.85}{\micro\meter\per\second}$ and below $\SI{0.35}{\micro\meter\per\second}$. The red curves correspond to siCTL cells, green to siITGB1 cells, and blue to siCD44 cells. For each figure, the straight lines correspond to the linear regressions of the weighted means of velocities after adjustment of the initial time in order to synchronize the oscillations. 
The data are normalized by $\SIlist{100;200;400}{\micro\meter\per\second}$. We normalize the data in order to facilitate the comparison between the different cohorts. The values considered will be explained later in Section~\ref{subsec:param_estimation_results}. However, it should be noted that the ratio between the selected velocities is equal to the ratio of the  pressure gradients.

\section{Methods}\label{sec:methods}
This study aims at deciphering the influence of hydrodynamic and adhesion forces on the dynamics of CTCs moving in interaction with the wall of a microfluidic device. {The methods regarding the experiments (cell culture, protein depletion, microfluidic design, etc.) can be found in~\cite{osmani}}. First, we perform a brief statistical examination of the tracked cell velocities in Subsection~\ref{subsec:statisticalStudy}. Second, a fluid velocity model is presented in Subsection~\ref{subsec:fluid_velocity}. Third, a model for CTCs velocity under both hydrodynamic transport and adhesion to the wall is derived in Subsection~\ref{subsec:cells_velocity}. 
In Subsection~\ref{subsec:param_estimation}, we fit the model to the experimental data using a well-designed technique for parameter estimation. {Finally, in Subsection~\ref{subsec:digital_twin}, the calibrated model for the first experiment allows to build a digital twin of flowing cells in adhesion.}

All statistical analyses were performed with \texttt{R}. For the t-tests, we use the \texttt{t$\_$test} function of the library \texttt{rstatix}.

\subsection{Quick statistical analysis of the data} \label{subsec:statisticalStudy}
We perform a quick statistical analysis of these velocities between the different cohorts and subcohorts {in each experiment}. First, we run t-tests of the mean velocity values to determine if the differences between cohorts and subcohorts are significant. Second, we run linear regressions on the velocity values and use them to determine for which cohort and subcohort the observed decreases are significant.

\subsection{Fluid velocity modeling}\label{subsec:fluid_velocity}
In this subsection, we derive a model for the fluid dynamics in the microfluidic device. When the viscous effects of the fluid prevail over convection, the Navier-Stokes equations can be reduced to a Poiseuille equation. In that case, the flow shows a parabolic profile at each time, with a maximal velocity in the center of the channel decreasing to zero at the walls. In case of a time-independent pressure gradient, the Poiseuille regime is valid when the fluid verifies the following properties : 
\begin{enumerate} 
\item[(1)] it is incompressible and Newtonian ; 
\item[(2)] the gravitational effect on the fluid is negligible;
\item[(3)] its flow is laminar ; 
\item[(4)] and its velocity profile does not evolve over the pipe's length denoted by $L$ in what follows.
\end{enumerate}
Conditions~1 and~2 are allowed when working with a microfluidic device where the fluid is mainly comparable to water.
Condition~3, can be checked by calculating the Reynolds number given by 
\[Re= \frac{\rho\, Q \, D_h}{\mu\, S},\]
where $\rho$ is the fluid density, $Q$ the volumetric flow rate, $\mu$ the dynamic viscosity, $D_h$ = $\frac{2(l\times h)}{l+h}$ the hydraulic diameter of a fully submerged rectangular channel and $S = l\times h$ the cross-section surface.  
The density can be taken as $\rho
 = \SI{1.00d3}{\kilo \gram \per \cubic \meter}$. Based on the experiments and procedure of Osmani and coworkers in~\cite{osmani,osmani2}, we have $D_h 
 = \SI{7.24d-4}{\meter}$, 
 $S  = \SI{1.52d-6}{\square \meter}$ and $Q\leq \SI{5.67d-9}{\cubic \meter \per \second}$.
 For the dynamic viscosity, one may refer to~\cite{Poon2022} (Table 2) to obtain a close approximation of the value (different medium, but similar composition when no FBS is added). The corresponding value is $\mu = \SI{7.31d-4}{\pascal \second}.$

It follows that $Re \leq 3.75$. This value is much smaller than the critical Reynolds number for the transition from a laminar to a turbulent state, that is equal to $2600$ in the case of a rectangular tube with a width eight times larger than the height, see~\cite{kao}. 
Finally, in order to verify Condition~4, we must determine the hydrodynamic entrance length of our microfluidic device (for more details, see~\cite{Incropera1996}, Chapter~8, Section~8.1).
For rectangular channels at laminar flow, a formula has been derived in~\cite{Ferreira2021}.
The hydrodynamic entrance $\ell$ (in meters) is then given as a non-linear function of the aspect ratio $AR =\frac{h}{l}$ and the Reynolds number $Re$.
Using the upper bound of $Re$ determined previously, a quick computation gives $\ell \leq \SI{1.53d-3}{\meter}$.  We can thus consider the Poiseuille flow to be fully developed (\textit{e.g.} independent of the length $L$) if we observe at least $\SI{1.53d-3}{\meter}$ away from the pipe inlet. According to the protocol given in~\cite{osmani2} the data was collected as close as possible to the center of the device lengthwise, which is at about $\SI{8.50d-3}{\meter}$ meters away from the inlet. 

In our case, however, the hypothesis of a time-independent pressure gradient is not valid. In fact, the fluid dynamics is affected by the angular velocity of the pump rotor, leading to a time-dependent oscillatory perturbation term to the pressure gradient term, which we model as follows
\[
G (1 + \xi_f \cos(\omega_f t + \varphi)), 
\]
where $\xi_f$ is the multiplicative correction amplitude, $\omega_f$  the angular velocity, and $\varphi$ the cell-dependent phase shift. 

When working with an oscillating pressure gradient, the condition for the establishment of a parabolic velocity profile is strongly tied to the frequency of the oscillation relative to the viscosity of the fluid. Such relation is given through a dimensionless coefficient $\mathbf{Wo} = \frac{h}{2}\sqrt{\dfrac{\omega_f \rho}{\mu}}$ introduced by Womersley in~\cite{womersley1955method},  which has to be inferior to 1 when the mean value of the pressure gradient is zero. A non-zero mean value will however relax such constraint, and using the results from~\cite{ma} along with supplementary observations, one can readily show a parabolic profile is obtained in our case, see Remark~\ref{rk:simusappendix} and Supplementary Material~\ref{subappendix:womersley} for more details. The fluid velocity can therefore be written as 
\begin{equation}\label{eq:solution1dN-S}
	 \frac{{G}}{2\nu\rho} \height_f^m (\height-\height_f^m)+\xi_f \frac{G}{\rho\omega_f} i \left(\frac{\sinh(Z \height_f^m)+\sinh(Z(\height-\height_f^m))}{\sinh(Z\height)} - 1\right) e^{i ( \omega_f t +\varphi)},
\end{equation}
where $i$ is the imaginary unit and $Z= (1+i) \sqrt{\frac{\omega_f}{2\nu}}$ with $\nu=\frac{\mu}{\rho}$ the kinematic viscosity.  We recall that $\height_f^m$ is the distance between the wall and cells in the focal plane and $\height = \SI{4.00d-4}{\meter}$ is the channel height. Taking the real part of this solution and performing computations (based on the linearity of the system and the principle of superposition), the fluid velocity in our context writes 
\begin{equation}\label{eq:compact_fluid_eq}
	u_f(t) =  \bar{u}_f(\height_f^m) + i_f(\height_f^m,\xi_f,\omega_f) \cos(\omega_f t +\varphi) + r_f(\height_f^m,\xi_f,\omega_f) \sin(\omega_f t+\varphi),
\end{equation}
where 
\begin{align*}
    \bar{u}_f(\height_f^m) & = \frac{G}{2\nu\rho} \height_f^m(\height-\height_f^m), \\
    r_f(\height_f^m,\xi_f,\omega_f)  & = \frac{\xi_f G}{\omega_f\rho}  \text{Re} \left( 1- \frac{\sinh(Z\height_f^m)+\sinh(Z(\height-\height_f^m))}{\sinh(Z\height)}\right), \\
    i_f(\height_f^m,\xi_f,\omega_f) &  = \frac{\xi_f G}{\omega_f\rho}  \text{Im} \left( 1- \frac{\sinh(Z\height_f^m)+\sinh(Z(\height-\height_f^m))}{\sinh(Z\height)}\right).
\end{align*}
Note that $\bar{u}_f$ is the mean fluid velocity, while the unknown parameters are $\height_f^m$, $\xi_f$, $\omega_f$ and $\varphi$.

\begin{remark} \label{rk:simusappendix}
Numerical approximations for the full device were also performed to test the hypothesis and investigate its limitations. These results confirming our fluid modelling can be found in the Supplementary Materials~\ref{appendix:fluid}. More specifically, we begin by verifying the value of the hydrodynamic entrance length, see Subsection~\ref{subappendix:hydrolength}, and we follow with validation of the fluid expression in Subsection~\ref{subappendix:1Dfluid}. A final subsection~\ref{subappendix:womersley} focuses on how the non-zero mean value of the pressure gradient allows us to work with a Poiseuille velocity profile.
\end{remark}

\subsection{Cells velocity modeling}\label{subsec:cells_velocity}
In this subsection, we define a deterministic model for cell motion based on a coupling between fluid velocity and adhesion dynamics, following previous studies~\cite{grec,etchegaray}. 
The interest is in capturing the different behaviours induced by varying fluid velocity and the number of expressed proteins. Both changes have an impact on cells velocity.

By denoting $N$ the bonds density and $u_c$ the cell velocity, the model writes $\forall t > 0$,
\begin{equation}
\label{eq:ode}
	\left\{ 
	\begin{array}{ll}
	N'(t) = c + (r - d)  N,\\[2ex]
	u_c(t) = u_f(t) - B( u_f(t),  u_c(t) ) \, N(t),
	\end{array}
	\right.
\end{equation}
together with the initial condition $N(0)=0$. In System~\eqref{eq:ode}, $c,\,r,\,d$ are given in \unit{\per \second} and stand respectively for the global binding rate, the growth rate and the unbinding rate. The function $B$ accounts for the velocity decrease arising from a unit adhesion density. All parameters are nonnegative.

\begin{assump} \label{assump:hp1}
We assume that the adhesion parameters are time-independent and depend only on the mean fluid velocity:
\[
c=c(\bar{u}_f), \, r=r(\bar{u}_f), \text{ and } d=d(\bar{u}_f).
\]
This amounts to neglecting the effects of fluid velocity oscillations on the binding dynamics.
\end{assump}

The function $B$ can depend either on fluid velocity or on the cell one. Three models are considered:
\begin{enumerate}
    \item Constant force model: 
    $ 
        B( u_f(t),  u_c(t) )= b\,,
    $ 
    where $b$ (in $\si{\micro\meter\per\second}$) quantifies the absolute velocity decrease induced by each unit of bonds density.
    \item Fluid-dependent force model:  
    $
        B( u_f(t),  u_c(t) )= b\, u_f(t)\,,
    $  
    with $b$ is the dimensionless proportion of velocity decrease induced by each unit of bonds density.
    \item Cell-dependent force model: 
    $
        B(u_f(t), u_c(t)) = b \, u_c(t)\,,
    $
    where $b$ is the dimensionless parameter for the friction ratio between bonds stiffness and fluid viscosity.
\end{enumerate}
\vspace{3mm}

\begin{assump} \label{assump:hp2}
We consider the cell-dependent model for the cell velocity equation: 
\[
    B(u_f(t), u_c(t)) = b \, u_c(t).
\]
\end{assump}

\vspace{3mm}

Under Assumptions~\eqref{assump:hp1}-\eqref{assump:hp2}, System~\eqref{eq:ode} has an explicit solution. If $d-r \ne 0$, we obtain for $t>0$ 
\begin{equation}\label{eq:n}
	N(t) = \dfrac{c}{d-r}( 1-e^{-(d-r) t} ),
\end{equation}
and then
\begin{equation}\label{eq:u_c}
	u_c(t) = \frac{u_f(t) }{1 - \frac{b c}{d-r} (e^{-(d-r)t}-1)} .
\end{equation}
Note that at $u_c(0) = u_f(0)$, so that the cell has not formed adhesion bonds at initial time. 
On the other hand, experimental observations may occur only after the cell has initiated an adhesive interaction with the wall. This is why we introduce an additional parameter $\tau \geq 0$ that stands for the observation time lag. A cell with a small value of $\tau$ would be observed with an initial velocity approaching the fluid velocity, while a cell with a large value of $\tau$ would be entering the observation zone with a lower velocity.

Finally, the percentage of decrease between the cell and the fluid velocities at time $t\geq 0$ is given by the quantity $1-\frac{u_c}{u_f}(t)$. Its limit as $t \rightarrow \infty$ then quantifies the asymptotic cell regime, and is given by
$$d_\% := \frac{bc}{d-r+bc}.$$

To conclude, our coupled model of parameters $\height_f^m$, $\xi_f$, $\omega_f$, $\varphi$, $\tau$, $b$, $c$, $r$ and $d$ reads	
\begin{equation}\label{eq:u_c_final}
    \begin{array}{ll}
    u_f(t) & = \bar{u}_f(\height_f^m) + i_f(\height_f^m, \xi_f, \omega_f) \cos(\omega_f t+\varphi) + r_f(\height_f^m, \xi_f, \omega_f) \sin(\omega_f t +\varphi), \\[1.5ex]
	u_c(t) & = u_f(t) / (1 - \frac{b c}{d-r} (e^{-(d-r)(t+\tau)}-1)).
    \end{array}
\end{equation}

\subsection{Parameters estimation}\label{subsec:param_estimation}
We calibrate the model using a well-adapted estimation procedure. The main difficulties in fitting our model 
to the data 
are (a) the little information on the fluid velocity, (b) the data noise (see Figure~\ref{fig:vel_mean}), and (c) the fact that the adhesion parameters are strongly correlated with the fluid parameters. 
Therefore,  we choose a mixed-effects parameter estimation procedure. The nonlinear mixed-effects model consists of pooling all subjects in a population and estimating a global distribution of uncertainties in the population to compensate for identifiability problems~\cite{lavielle2014mixed}. For example, the parameters of each cell $i$ could be divided into two types of uncertainties:
a first part that is the same for all cells (denoted by $\theta_{pop}$ for the parameter $\theta^{i}$) and corresponds to the fixed effect, and 
a second part that represents individual variability (denoted by $\theta_{ind}^{i}$) and corresponds to random effects, \textit{i.e.} $\theta^{i} = \theta_{pop}+\theta_{ind}^{i}$. Different covariates can also be added, e.g., for different cohorts of the population. A nonlinear mixed-effect estimation algorithm --~called the stochastic approximation expectation maximization (SAEM) algorithm~\cite{kuhn}~-- is implemented in the software \texttt{Monolix}~\cite{monolix}. Thanks to the R package \texttt{lixoftConnectors}, we could easily run \texttt{Monolix} using R. The code and extracted cell velocities are available \href{https://plmlab.math.cnrs.fr/gciavolella/ctc_adhesion_microfluidic}{here} (a recent version of \texttt{Monolix} is required).

Since the mean fluid velocity is too high for the tracker to automatically track the CTC as the highest pressure gradient, we do not track the cells in all the 3 experiments but only for Experiment~1. However, it is useful to have a case with 3 velocities to make the estimation procedure more robust. This motivates us to propose an estimation strategy that is decomposed into two parts:
\begin{itemize}
\item[(1)] {Cell modeling choice and fluid reconstruction with} estimation of the fluid parameters (${\height_f^m}$, $\omega_f$, $\xi_f$), the adhesion parameters ($bc$, $d-r$) and the time parameters ($\phi$ and $\tau$) using Experiment~$1$, as this is the configuration with the most available data: 3 velocities were tracked and 3 different cells (siCTL, siITGB1, siCD44) ;
\item[(2)] Estimation of only one fluid parameter (${\height_f^m}$ corresponding to the distance between the wall and the cells in the focal plane) and the adhesion parameters ($bc$, $d-r$) as well as the time parameters ($\phi$ and $\tau$) using all experiments.
\end{itemize}
In this second step, the other fluid parameters are set to the values obtained in the first step and the parameter priors were set to the estimated values in the first step. See the following two subsections for more details on each of these steps.

\subsubsection{{Fluid parameters estimations thanks to Experiment 1}} \label{ref:estimparam1st}

In this first step, we recall that the data of Experiment~$1$ are used (137 trajectories). 

\paragraph{Fluid parameters}
The only known fluid parameter is the mean pressure gradient given by $G^{(\coh)} = 2^{\coh-1}G^{(1)}$, where $\cdot^{(\coh)}$  denotes here and in the following the velocity cohort for $\coh \in \{1, \; 2, \; 3\}$, and $G^{(1)}=\SI{50.33}{\pascal\per \meter}$. 
{The fluid setting is going to be the same for all cells meaning that $\height_f^m$, $\omega_f$ and $\xi_f$ are estimated without considering individual variability.}
In addition always to avoid identification problems, we strongly incorporate the information we have between the three velocity cohorts by considering
\[ 
	{\height_f^m}^{(\coh)} = {\height_f^m}^{(1)} \; \text{m}, 	\quad  \omega_f^{(\coh)} = 2^{\coh-1} \,  \omega_f^{(1)} \; \text{rad.s}^{-1}, 
		\quad \xi_f^{(\coh)} =  2^{1-\coh}  \xi_f^{(1)}, 
\]
for $\coh \in \{1,2,3\}$.
The hypothesis on $\omega_f$ is obvious and the hypothesis on $\xi_f$ implies that the oscillation amplitude of the pressure gradient is constant as $G^{(\coh)} \xi_f^{(\coh)} = G^{(1)} \xi_f^{(1)}$.  Only the phase shift $\varphi$ depends on the considered cell.

This implies that $142$ parameters only for fluid velocity must be estimated: $3$ fixed effects for ${\height_f^m}^{(1)}, \omega_f^{(1)}, \xi_f^{(1)}$, $2$ fixed effects for~$\varphi$ (mean and standard deviation) and $137$ randoms effects for~$\varphi$.

\paragraph{Adhesion parameters}
Since $b$, $c$, $r$, and $d$ are strongly paired, they can not be identified independently from the observations. We will then estimate only $bc$ and $d-r$. 

The fact that we use a mixed effect approach on Experiment~$1$ has the great advantage that we can constrain the values of the fluid parameters, but it introduces a bias in the estimation of the individual variability of the parameters, since the fixed effects are the same for the 3 velocity cohorts and for the 3 types of cells. It is then possible to consider cohort covariates and here we decide to consider
\[\log(bc^i) = \log(bc_{pop}) + \beta^{(2)}_{bc}[\text{if } \coh=2] + \beta^{(3)}_{bc}[\text{if } \coh=3] + bc^i_{ind}.\]

For the adhesion parameters $bc$, $d-r$ and $\tau$, the behaviour of the individual cell is integrated meaning that the number of parameters to be estimated is $420$: $3\times 2$ fixed effects (mean and std), $3$ covariates for $bc$ and $3 \times 137$ random effects. 

\paragraph{Distribution laws} 
For parameter distributions, we consider \textit{logitnormal} distributions for $\height_f^m$, $\xi_f^m$, $\varphi$ and $\tau$ to keep them respectively in
\SIrange{0}{1.5 d-5}{\meter}, \numrange{0}{1}, \numrange[parse-numbers = false]{0}{2\pi}, and  \SIrange{0}{10}{\second}.  
We consider \textit{lognormal} distributions for the other parameters. Furthermore, whenever mixed effects are estimated, prior values and prior standard deviations are given to the SAEM algorithm. We will denote by the superscript $(\cdot)^*$ the priors and by $(\cdot)^s$ the prior standard deviations. We consider
\begin{equation*} ({\height_f^m}^{(1)})^* = \SI{7.5d-6}{\meter}, \; (\omega_f^{(1)})^* = 12\; \text{rad.s}^{-1}, \; (\xi_f^{(1)})^* = 0.3,  \; (\varphi)^* = \pi, \; (\varphi)^s = 1, 
\end{equation*}
\begin{equation*}
(bc)^* = 0.5\; \text{s}^{-1}, \; (bc)^s =1, \; (d-r)^* = 0.5\;\text{s}^{-1}, \; (d-r)^s =1 \; (\tau)^* = 5 \;\text{s}, \; (\tau)^s = 1.
\end{equation*}

\paragraph{Model error} 
We consider a constant error model and we estimate the  error model standard deviation starting from the prior value~$25$. Adding this value to the estimation of the fluid and adhesion parameters, 563 parameters are estimated during the procedure.  

\subsubsection{{Estimation of parameters coupling all experiments}} \label{ref:estimparam2nd}

To facilitate comparison with the first estimation, we only indicate the differences:
\begin{itemize}
	\item all the trajectory data are used,
	\item $\omega_f^{(1)}$ and $\xi_f^{(1)}$ are fixed with the estimated parameters of the first step,
	\item a covariate corresponding to the experiment is added for the estimation of  ${\height_f^m}^{(1)}$ such that the value for Experiment~$1$ is fixed to the already estimated value,
	\item the priors values of ${\height_f^m}^{(1)}$, $bc$, $d-r$, $\varphi$ and $\tau$ are set to the mean estimates of the first step,
	\item and there is no random effect on $d-r$.
\end{itemize}
This second step leads to the estimation of 720 parameters: $2$ covariates for ${\height_f^m}^{(1)}$ and $3$ for $bc$, $1$ fixed effect for $d-r$, $3\times 2$ fixed effects (mean and standard deviation) and $3\times 236$ random effects for $bc$, $\varphi$ and $\tau$.
	
\subsection{Digital twin}\label{subsec:digital_twin}

{The fact that our approach is based on a mathematical model allows the reproduction of a digital twin of the experimental setup --~ once the parameters are estimated~-- even for {fluid} configurations that were not considered during the experiments. }{The procedure has been tested for Experiment~$1$. For that purpose, we use the reconstructed fluid model at a chosen pressure gradient,  set $\tau,\,  \varphi=0$, and consider the estimated values of the adhesion parameters to simulate synthetic cell velocities in each subcohort for $2000$ cells.  Parameters with individual variability are assumed to follow a Gaussian distribution with estimated mean and standard deviation. This allows to} {mathematically reconstruct the velocity behavior of the cells over time from their entry into the microfluidic channel. The limitation of the experiments to videos from a specific area can thus be overcome. }

We can go further by {building} a phenomenological law for parameters with individual variability as a function of the fluid velocity. By considering a linear interpolation {and using the estimated standard deviations of these parameters} we can generate synthetic cells in unobserved velocity configurations and obtain statistical quantities about the digital cohort. 

\section{Results}\label{sec:results}

\subsection{Quick statistical analysis of the data}

The \textit{p-values} resulting from the t-tests are given in Table~\ref{table:pvalues}. The diagonal blocks correspond to comparison between different protein modifications at the same fluid pressure, while the upper diagonal blocks account for comparison of subcohorts having the same protein modification and different fluid pressures. Significant differences can be seen between different protein modifications at the same fluid pressure and between the same protein modifications at different fluid pressure.
Indeed, \textit{p-values} are smaller than $4\times 10^{-3}$, except between between siITGB1$^{(3)}$ and siCD44$^{(3)}$ {($p = 0.6$)} and siITGB3$^{(1)}$ and siITGB3$^{(2)}$ {($p = 0.2$)}. Thus, at high fluid pressure gradients, the differences depleting the first or second adhesion protein are not relevant, but at smaller pressure gradients they are informative.

The outputs of the linear regressions can be found in Table~\ref{table:linear_regression}. 
The intercepts --~corresponding to the value of cell velocity at time $t=0$~-- increase with fluid pressure for the same protein modification, and for a given fluid pressure it increases with respect to protein modifications or it remains stable. Anyways, intercepts are always smaller than $1$, value corresponding to the normalisation by $\SIlist{100;200;400}{\micro\meter\per\second}$. 
The slope estimate and its \textit{p-value} --~related to the adhesions effects during the observation duration~-- show a significant decrease from the intercept value for almost all cases except for {siCTL$^{(1)}$ and siITGA5$^{(2)}$ in Experiment $2$ and siCTL$^{(1)}$ and siCTL$^{(2)}$ in Experiment $3$}.

\captionsetup[table]{labelfont=bf}
\begin{table}
	\begin{subtable}{\textwidth}
		\caption{\textbf{Experiment $\mathbf{1}$}}
		\scalebox{0.75}{
		\begin{tabular}{|c||c|c||c|c|c||c|c|c|c|} 
			\hline
			 & {\textbf{siITGB1$^{(1)}$}} & {\textbf{siCD44$^{(1)}$}} & {\textbf{siCTL$^{(2)}$}} & {\textbf{siITGB1$^{(2)}$}} & {\textbf{siCD44$^{(2)}$}} & {\textbf{siCTL$^{(3)}$}} & {\textbf{siITGB1$^{(3)}$}} & {\textbf{siCD44$^{(3)}$}}\Tstrut\Bstrut\\ 
			\hline\hline
			{\textbf{siCTL$^{(1)}$}}   & $\mathbf{< 10^{-7}}$ & $\mathbf{< 10^{-7}}$ & $\mathbf{< 10^{-7}}$ & - & - & $\mathbf{< 10^{-7}}$ & - & - \Tstrut\\[1ex] 
			{\textbf{siITGB1$^{(1)}$}} & - & $\mathbf{2.5\times 10^{-3}}$ & - &  $\mathbf{1.3 \times 10^{-5}}$ &    -  &   -   & $\mathbf{< 10^{-7}}$  &   - \\[1ex] 
			{\textbf{siCD44$^{(1)}$}}  &       -      &       -     &     -      &       -  &  $\mathbf{3.2 \times 10^{-3}}$ &    -   &  -   &  $\mathbf{< 10^{-7}}$ \\[1ex]  
			\hline \hline
			\rule{0pt}{3ex}  {\textbf{siCTL$^{(2)}$}}    &         -      &    -    &         -   &   $\mathbf{< 10^{-7}}$ &        $\mathbf{< 10^{-7}}$  &  $\mathbf{< 10^{-7}}$  &  -  &   - \\[1ex] 
			{\textbf{siITGB1$^{(2)}$}}    &         -   &       -      &       -        &    -      &       $\mathbf{5.3 \times 10^{-4}}$    &     -  & $\mathbf{< 10^{-7}}$  &   -      \\[1ex] 
			{\textbf{siCD44$^{(2)}$}}    &          -   &       -    &         -     &       -     &        -   &       -    &      -  & $\mathbf{< 10^{-7}}$  \\[1ex]\hline \hline
			\rule{0pt}{3ex} {\textbf{siCTL$^{(3)}$}}   &          -  &        -       &      -       &     -     &        -    &      -     &   $\mathbf{4.1 \times 10^{-6}}$   &     $\mathbf{< 10^{-7}}$ \\[1ex] 
			{\textbf{siITGB1$^{(3)}$}}    &          -  &        -       &      -       &     -     &        -    &      -     &     -  &  $0.6$ \\[1ex] 
			\hline
		\end{tabular}}
		\centering
		\caption{\textbf{Experiment $\mathbf{2}$}}
		\scalebox{0.9}{
		\begin{tabular}{|c||c||c|c|} 
			\hline
			 & {\textbf{siITGA5$^{(1)}$}}  & {\textbf{siCTL$^{(2)}$}} & {\textbf{siITGA5$^{(2)}$}} \Tstrut\Bstrut\\ 
			\hline\hline
			{\textbf{siCTL$^{(1)}$}}   &  $\mathbf{< 10^{-7}}$& $\mathbf{3.3\times 10^{-3}}$   & - \Tstrut\\[1ex] 
			{\textbf{siITGA5$^{(1)}$}} & -  & - & $\mathbf{1.2 \times 10^{-4}}$ \\[1ex] 
			\hline \hline
			\rule{0pt}{3ex}  {\textbf{siCTL$^{(2)}$}}    &         -      &  - &    $\mathbf{< 10^{-7}}$    \\[1ex] 
			\hline
	\end{tabular}}
	\centering
	\caption{\textbf{Experiment $\mathbf{3}$}}
	\scalebox{0.9}{
		\begin{tabular}{|c||c||c|c|} 
			\hline
			&{\textbf{siITGB3$^{(1)}$}}  & {\textbf{siCTL$^{(2)}$}} & {\textbf{siITGB3$^{(2)}$}} \Tstrut\Bstrut\\ 
			\hline\hline
			{\textbf{siCTL$^{(1)}$}}   & $\mathbf{6.9 \times 10^{-5}}$  & $\mathbf{6.4 \times 10^{-5}}$  & - \Tstrut\\[1ex] 
			{\textbf{siITGB3$^{(1)}$}} & - & - & ${0.2 }$ \\[1ex] 
			\hline \hline
			\rule{0pt}{3ex}  {\textbf{siCTL$^{(2)}$}}    &    -     &  -   & $\mathbf{9.8 \times 10^{-6}}$  \\[1ex] 
			\hline
	\end{tabular}}
	\end{subtable}
	\caption{Resulting \textit{p-values} from the \textit{t-tests} performed on the mean velocity values between different protein modifications at the same fluid pressure and between the same protein modification but at different fluid pressures. Superscripts refer to the experiment number with the fluid pressure values as in Subsection~\ref{subsec:data_availability}. The bold values are the \textit{p-values} inferior to {$4\times 10^{-3}$}.}
	\label{table:pvalues}
\end{table}

\begin{table}
	\begin{subtable}{\textwidth}
		\centering
		\caption{\textbf{Experiment $\mathbf{1}$}}
		\begin{tabular}{|c|c|c|c|c|} 
			\hline
			\backslashbox{Experiment}{Regression}& Intercept estimate & Slope estimate & Slope \textit{p-value} \\ 
			\hline
			{\textbf{siCTL$^{(1)}$}}  & $0.49$ & $-0.006 $ & {$\mathbf{< 10^{-5}}$} \\
			{\textbf{siITGB1$^{(1)}$}} & $0.57$ & $-0.007$ & {$\mathbf{< 10^{-5}}$}\\
			{\textbf{siCD44$^{(1)}$}} & $0.59$ & $-0.0035$ & {$\mathbf{1.5 \times10^{-2}}$} \\
			{\textbf{siCTL$^{(2)}$}} & $0.54$ & $-0.01$   & {$\mathbf{1.6\times 10^{-5}}$} \\
			{\textbf{siITGB1$^{(2)}$}} & $ 0.62$  & $-0.02$   & {$\mathbf{< 10^{-5}}$}\\
			{\textbf{siCD44$^{(2)}$}} & $0.61$  &  $-0.009$ &  {$\mathbf{4.9\times 10^{-2}}$}  \\
			{\textbf{siCTL$^{(3)}$}} & $0.65$  &  $-0.027$    & {$\mathbf{1.1\times 10^{-3}}$}\\
			{\textbf{siITGB1$^{(3)}$}} & $0.71$  & $-0.054$   &  {$\mathbf{< 10^{-5}}$} \\
			{\textbf{siCD44$^{(3)}$}} & $0.70$   &  $-0.053$ & {$\mathbf{< 10^{-5}}$} \\
			\hline
		\end{tabular}
	\end{subtable}
	\begin{subtable}{\textwidth}
		\centering
		\caption{\textbf{Experiment $\mathbf{2}$}}
		\begin{tabular}{|c|c|c|c|} 
			\hline
			\backslashbox{Experiment}{Regression}& Intercept estimate & Slope estimate & Slope \textit{p-value} \\ 
			\hline
			{\textbf{siCTL$^{(1)}$}}  & $0.43$ & $-0.002 $ & {${2.4 \times 10^{-1}}$} \\
			{\textbf{siITGA5$^{(1)}$}} & $0.52$ & $-0.006$ & {$\mathbf{2.9\times 10^{-3}}$}\\
			{\textbf{siCTL$^{(2)}$}}  & $0.46$ & $-0.011 $ & {$\mathbf{1.8 \times 10^{-2}}$} \\
			{\textbf{siITGA5$^{(2)}$}} & $0.57$ & $-0.0016$ & {${7\times 10^{-1}}$}\\
			\hline
		\end{tabular}
	\end{subtable}
	\begin{subtable}{\textwidth}
		\centering
		\caption{\textbf{Experiment $\mathbf{3}$}}
		\begin{tabular}{|c|c|c|c|} 
			\hline
			\backslashbox{Experiment}{Regression}& Intercept estimate & Slope estimate & Slope \textit{p-value} \\ 
			\hline
			{\textbf{siCTL$^{(1)}$}}  & $0.42$ & $-0.0005 $ & {${7.2 \times 10^{-1}}$} \\
			{\textbf{siITGB3$^{(1)}$}} & $0.54$ & $-0.014$ & {$\mathbf{< 10^{-5}}$}\\
			{\textbf{siCTL$^{(2)}$}}  & $0.45$ & $0.008 $ & {${1.0 \times 10^{-1}}$} \\
			{\textbf{siITGB3$^{(2)}$}} & $0.52$ & $-0.014$ & {$\mathbf{< 10^{-5}}$}\\
			\hline
		\end{tabular}
	\end{subtable}
	\caption{ Linear regressions of the velocity values for all cohorts and subcohorts. Column~2: Intercept estimates (cell velocity value in the regression when $t=0$).  Column~3: Slope estimates. Column~4: \textit{p-value} of the slope estimates. Superscripts refer to the experiment number with the fluid pressure values as in Subsection~\ref{subsec:data_availability}. The bold values are the \textit{p-values} inferior to $5\times 10^{-2}$.}
	\label{table:linear_regression}
\end{table}

\subsection{Parameters estimation}\label{subsec:param_estimation_results}
\paragraph{Fluid parameters}
In what follows, when the parameters depend only on the velocity cohort, the 3 estimated values are given in a vector where the $\coh^{th}$ value corresponds to the $\coh^{th}$ cohort for $\coh \in \{1,2,3\}$. 
The results of the first estimation (using the 3 velocities but only Experiment~1, see Subsection~\ref{ref:estimparam1st} for details) allows to estimate the following values for the fluid parameters
\[
\quad \xi_f^{(\coh)} =(0.27, \, 0.13, \, 0.07),  \quad \omega_f^{(\coh)} = (12.2,\, 24.4 ,\, 48.8)\;\text{rad.s}^{-1}, 
\] 
and ${\height_f^m}^{(1)} = \SI{7.2d-6}{\meter}$ corresponding them to Experiment~$1$. Using the second estimation, we obtain for ${\height_f^m}^{(1)} = \SI{6.8d-6}{\meter}$ for Experiment~$2$ and ${\height_f^m}^{(1)} = \SI{6.0d-6}{\meter}$ for Experiment~$3$. 

Using these estimated values, the mean velocity values can be computed using the first term of Equation~\eqref{eq:solution1dN-S} and we obtain
\begin{align*}
	\bar{u}_f^{(\coh)} = & (99.25, \, 198.5, \,   397.0) \; \SI{}{\micro\meter\per \second} \\
	 \text{(respectively } \bar{u}_f^{(\coh)} = & (92.9, \, 185.7, \,   371.4) \;
\SI{}{\micro\meter\per \second} \text{ and } \bar{u}_f^{(\coh)} = (83.2, \, 166.4, \,   332.8) \; \SI{}{\micro\meter\per \second})
\end{align*}
for Experiment~1 (respectively for Experiment~2 and Experiment~3).

\begin{remark}
These estimated mean velocity values justify the normalization considered in Figure~\ref{fig:vel_mean}.
\end{remark}

\begin{figure}
\centering
\begin{tabular}{ccc}
\includegraphics[width=0.8\textwidth,trim = {0cm 0cm 0cm 0cm},clip]{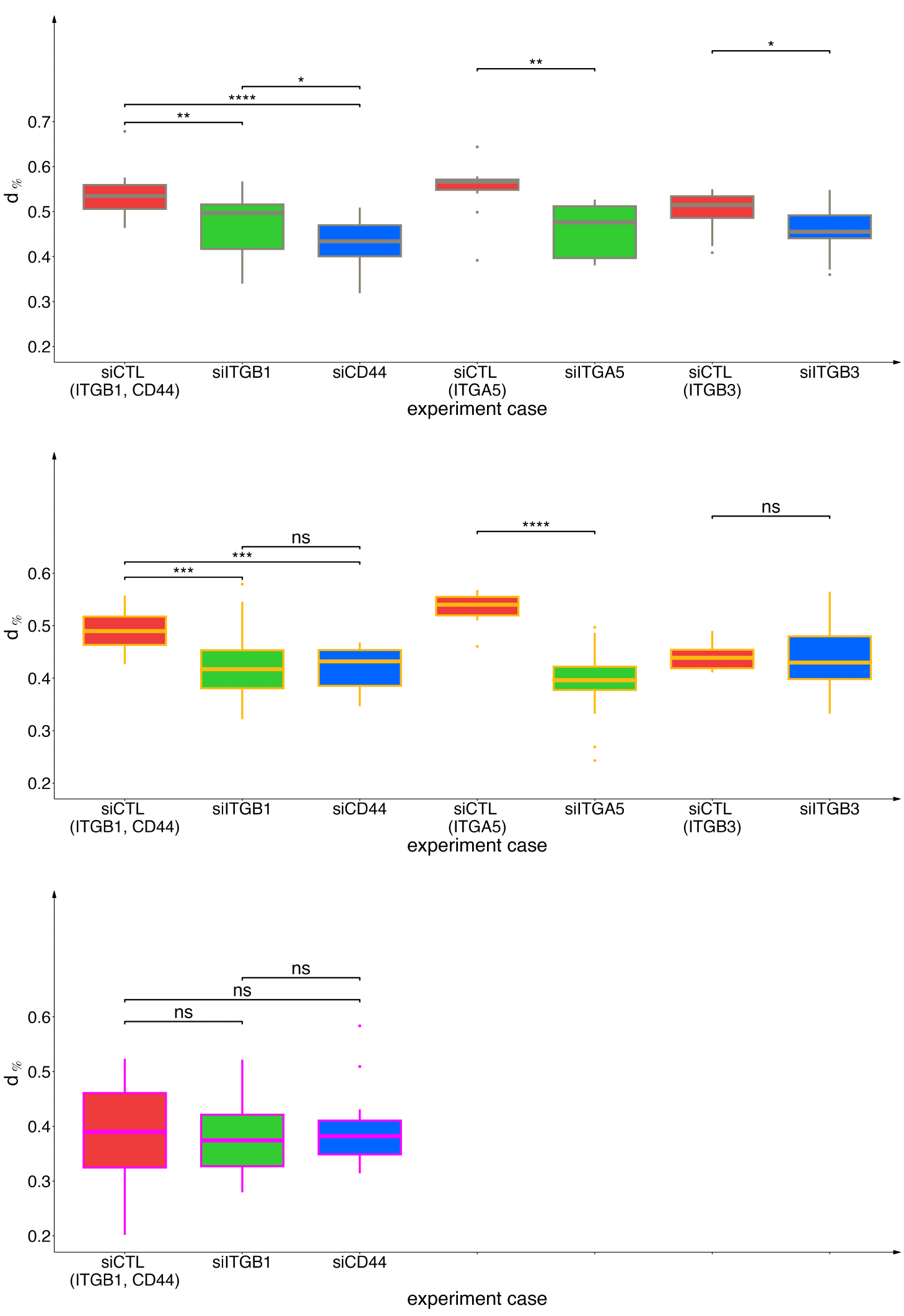} \\[1ex]
\includegraphics[width=0.6\textwidth]{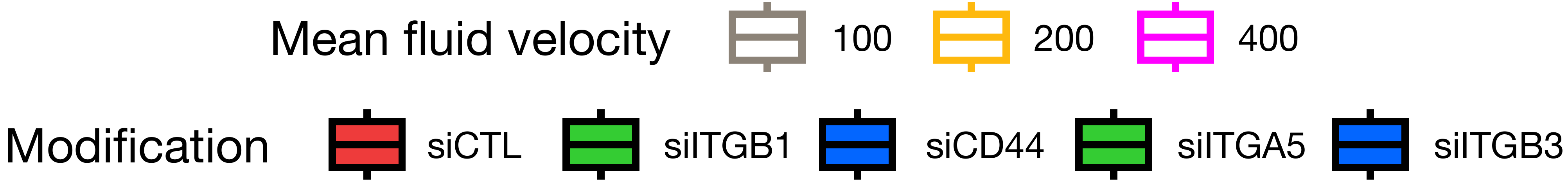}
\end{tabular}
\caption{Box plots of the percentage of decrease in cell velocity $d_\%$ for the 3 experiments (red: siCTL for the three experiments, green: siITGB1 and siITGA5, blue: siCD44 and siITGB3) and for the 3 pressure gradient cohorts (top: pressure gradient fixed at $G^{(1)}$ corresponding to $\sim\SI{100}{\micro\meter\per\second}$, middle: $G^{(2)}$ corresponding to $\sim\SI{200}{\micro\meter\per\second}$, bottom: $G^{(3)}$ corresponding to $\sim\SI{400}{\micro\meter\per\second}$). We substitute p-values with symbols such that: ns correspond to p$>10\%$, $^{*}$ to p$\leq 10\%$, $^{**}$ to p$\leq 5\%$, $^{***}$ to p$\leq 0.5\%$, and $^{****}$ to p$\leq 0.05\%$. }
\label{fig:100200400modif}
\end{figure}

\begin{figure}
	\begin{center}	
\includegraphics[width=0.8\textwidth,trim = {0cm 0cm 0cm 0cm},clip]{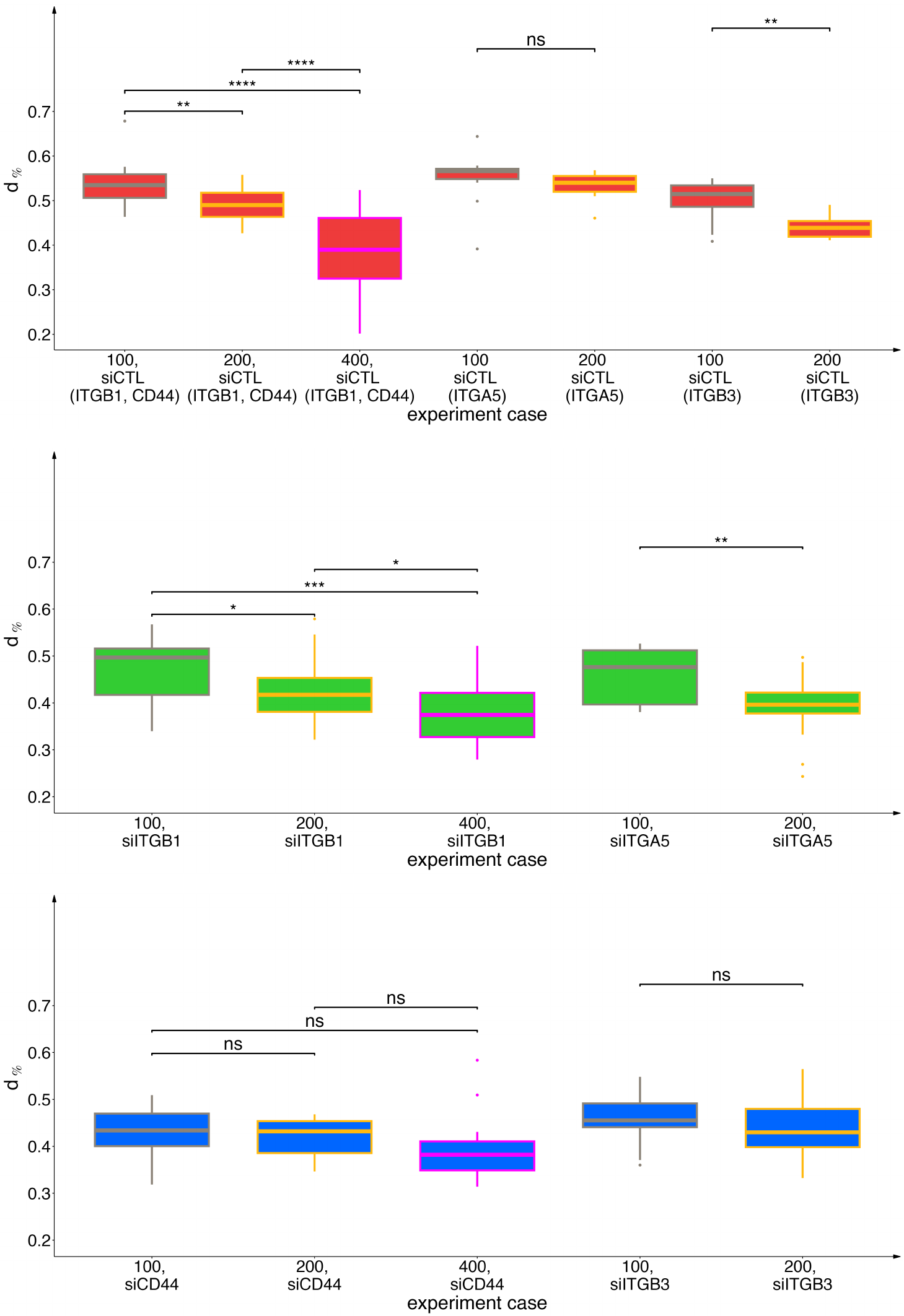} \\[1ex]
\includegraphics[width=0.6\textwidth]{legend.pdf}
\caption{Box plots of the percentage of decrease in cell velocity $d_\%$ for the 3 experiments and for the 3 pressure gradient cohorts (top: siCTL, middle: siCD44 and siITGB3, bottom: siITGB1 and siITGA5). We substitute p-values with symbols such that: ns correspond to p$>10\%$, $^{*}$ to p$\leq 10\%$, $^{**}$ to p$\leq 5\%$, $^{***}$ to p$\leq 0.5\%$, and $^{****}$ to p$\leq 0.05\%$. }
\label{fig:MODIFvelocity}
\end{center}
\end{figure}

For the adhesion parameters, we give the results obtained at the second estimation. The parameter $d-r$ is estimated at $0.62$ and {Figure}~\ref{fig:100200400modif} summarizes the individual estimated values. More precisely, it corresponds to the box plots of the percentage of decrease in cell velocity $d_\%$ resulting estimating $bc$ and $d-r$.
We also add \textit{p-values} significance ranges of the t-tests between the estimated values of parameters for different protein modifications at the same fluid pressure. To facilitate the comparison between the same protein modifications at different fluid pressure, Figure~\ref{fig:MODIFvelocity} shows the same estimated values but sorted by protein modifications instead of fluid pressure gradient. The \textit{p-values} ranges of the t-tests between the estimated values of parameters at different fluid pressure gradients for the same proteins are shown. 

\begin{figure}
	\begin{center}
		\includegraphics[width=.7\textwidth]{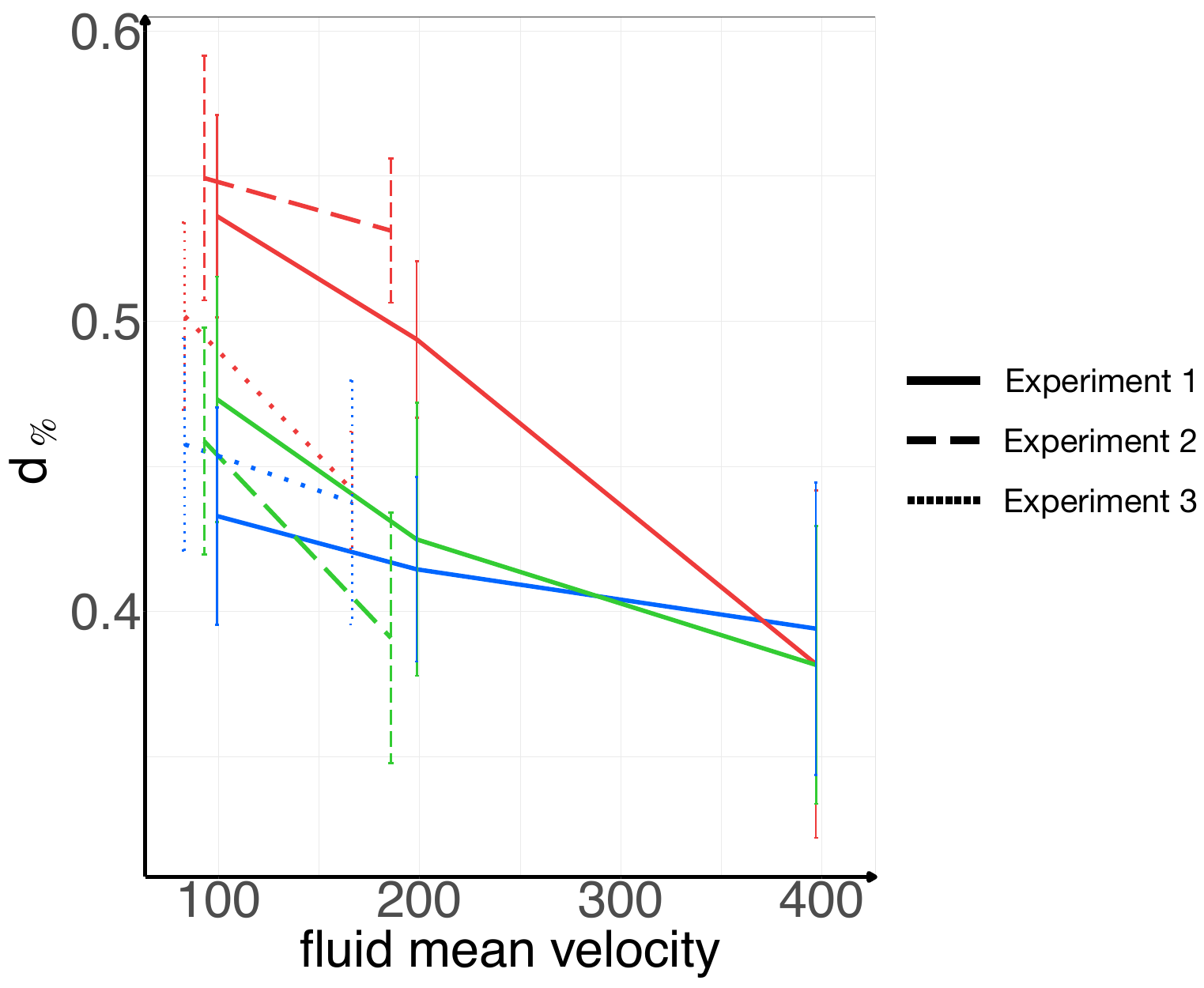} 
		\caption{Representation of the mean values of $d_\%$ for the different mean fluid velocities (x-axis, {given in micrometers per seconds}) and for the different protein modifications in Experiment~$1$ (solid lines), Experiment $2$ (dashed line), Experiment $3$ (dotted lines). The error bars correspond to the standard deviations.}
		\label{fig:mean}
	\end{center}
\end{figure}

Table~\ref{tab:meanpop} summarises the mean and standard deviations of $d\%$ by cohorts and subcohorts for each experiment. To facilitate the reading of this table, the mean values and the associated standard deviations are plotted in Figure~\ref{fig:mean}.

\begin{figure}
	\begin{center}	
		\includegraphics[width=\textwidth,trim = {0cm 0cm 0cm 0cm},clip]{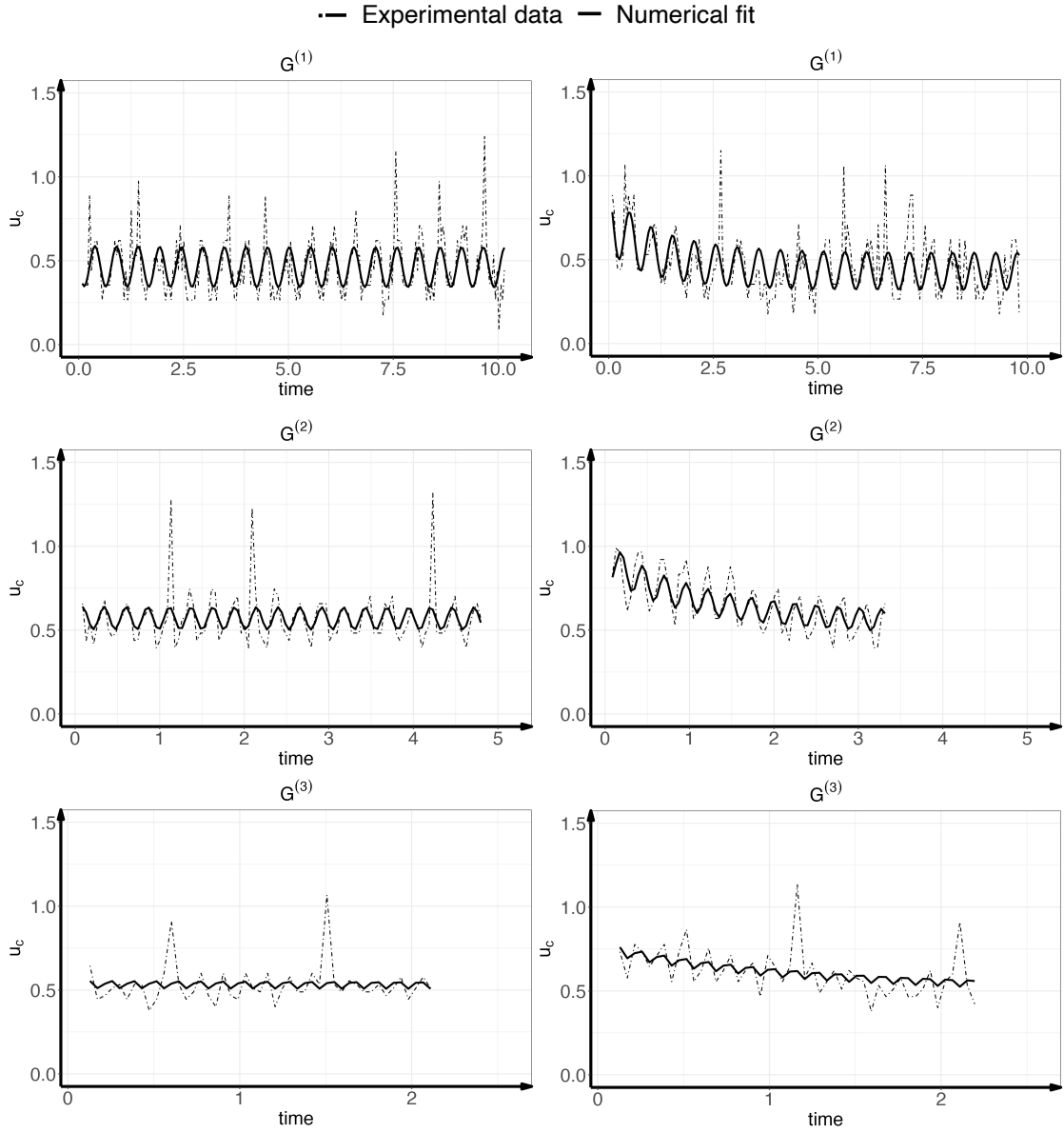}
		\caption{ Numerical fits for Experiment 1. Time is given in seconds, and velocities in micrometers per seconds. The dot-dash line represents the experimental cells velocity and in solid, the fit that we obtain with Equation~\eqref{eq:u_c_final}. Top: Pressure gradient  fixed at $G^{(1)}$ corresponding to $\sim\SI{100}{\micro\meter\per\second}$, middle: $G^{(2)}$ corresponding to $\sim\SI{200}{\micro\meter\per\second}$, bottom: $G^{(3)}$ corresponding to $\sim\SI{400}{\micro\meter\per\second}$. Left: CTCs with a stationary regime, right: CTCs with a decreasing velocity. The time interval decreases since CTCs observation duration reduces with higher velocity.}
		\label{fig:fit}
	\end{center}
\end{figure}

Finally, Figure~\ref{fig:fit} shows numerical fits for Experiment~$1$ compared to the experimental data. Independently of the fluid cohort, two typical behaviours are observed: the CTCs velocity either remains stationary or decreases. Therefore, we show examples of these behaviours for each velocity cohort in the siCTL case only. Velocity values are normalised by $2^{\coh-1} \times \SI{100}{\micro\meter\per\second}$ for each $\coh \in \{1,2,3\}$.

\captionsetup[table]{labelfont=bf}
\begin{table}
	\begin{subtable}{\textwidth}
		\centering
		\caption{\textbf{Experiment $\mathbf{1}$}}
	\begin{tabular}{|c|c|} 
		\hline
		\backslashbox{Experiment}{Values}& $d\%$ \\ 
		\hline
		{\textbf{siCTL$^{(1)}$}}  & 0.54 (0.05)   \\
		{\textbf{siITGB1$^{(1)}$}} & 0.47 (0.06)   \\
		{\textbf{siCD44$^{(1)}$}} & 0.43 (0.05)   \\
		{\textbf{siCTL$^{(2)}$}} & 0.49 (0.04)  \\
		{\textbf{siITGB1$^{(2)}$}} & 0.42 (0.07)   \\
		{\textbf{siCD44$^{(2)}$}} & 0.41 (0.05)   \\
		{\textbf{siCTL$^{(3)}$}} & 0.38 (0.09)   \\
		{\textbf{siITGB1$^{(3)}$}} & 0.38 (0.07)   \\
		{\textbf{siCD44$^{(3)}$}} & 0.39 (0.07)   \\
		\hline
	\end{tabular}
	\end{subtable}
	\begin{subtable}{\textwidth}
		\centering
		\caption{\textbf{Experiment $\mathbf{2}$}}
		\begin{tabular}{|c|c|} 
			\hline
			\backslashbox{Experiment}{Values}& $d\%$ \\ 
			\hline
			{\textbf{siCTL$^{(1)}$}}  & 0.55 (0.06)   \\
			{\textbf{siITGA5$^{(1)}$}} & 0.46 (0.06)   \\
			{\textbf{siCTL$^{(2)}$}} & 0.53 (0.04)  \\
			{\textbf{siITGA5$^{(2)}$}} & 0.39 (0.06)   \\
			\hline
		\end{tabular}
	\end{subtable}
	\begin{subtable}{\textwidth}
		\centering
		\caption{\textbf{Experiment $\mathbf{2}$}}
		\begin{tabular}{|c|c|} 
			\hline
			\backslashbox{Experiment}{Values}& $d\%$ \\ 
			\hline
			{\textbf{siCTL$^{(1)}$}}  & 0.50 (0.05)   \\
			{\textbf{siITGB3$^{(1)}$}} & 0.46 (0.05)   \\
			{\textbf{siCTL$^{(2)}$}} & 0.44 (0.03)  \\
			{\textbf{siITGB3$^{(2)}$}} & 0.44 (0.06)   \\
			\hline
		\end{tabular}
	\end{subtable}
	\caption{Mean and standard deviation values of $d\%$ for the all cohorts and subcohorts. The first value corresponds to the mean and the second value in parenthesis to the standard deviation. }
	\label{tab:meanpop}
\end{table}

\subsection{Digital twin}
\begin{figure}
	\begin{center}
		\includegraphics[width=.9\textwidth]{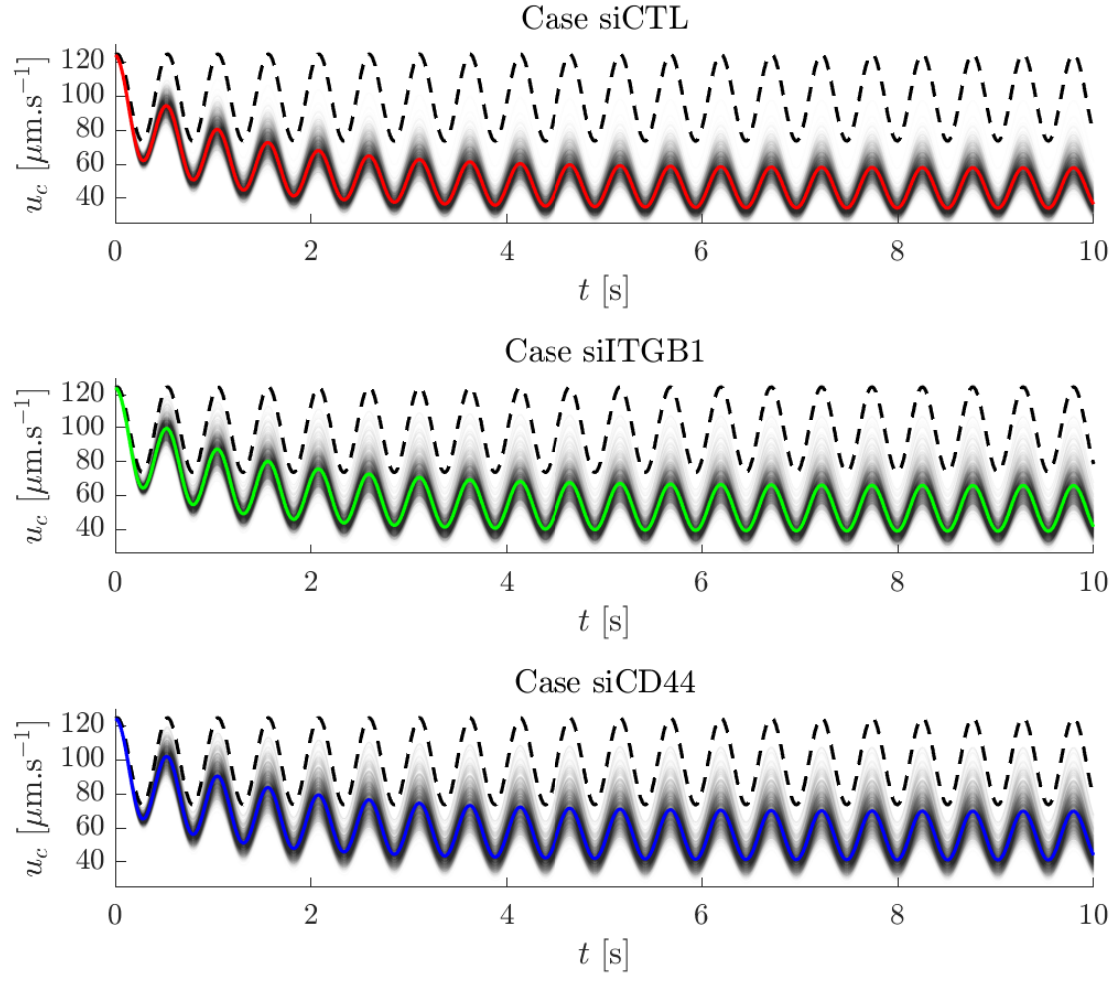}
		\caption{{Generation of $2000$ synthetic cells when $G = G^{(1)} = \SI{50.33}{\pascal \per \meter}$ for the $3$ protein cases {of Experiment 1} with $d-r = 0.62$, $bc \sim \mathcal{N}(m_{bc},\sigma_{bc})$, where $m_{bc}$ and $\sigma_{bc}$ are the estimated values summarized in Table~\ref{tab:meanpop}. In opaque black their velocities $u_c$ over $10 s$, calculated from Model~\eqref{eq:u_c_final} with $\tau=0$, in red (siCTL), green (siITGB1) and blue (siCD44) the evaluated mean value over time and in dashed black the fluid velocity $u_f$ in Equation~\eqref{eq:compact_fluid_eq} with $\phi=0$.}}
		\label{fig:cellspredict}
	\end{center}
\end{figure}

\begin{figure}
	\begin{center}
		\includegraphics[width=.9\textwidth]{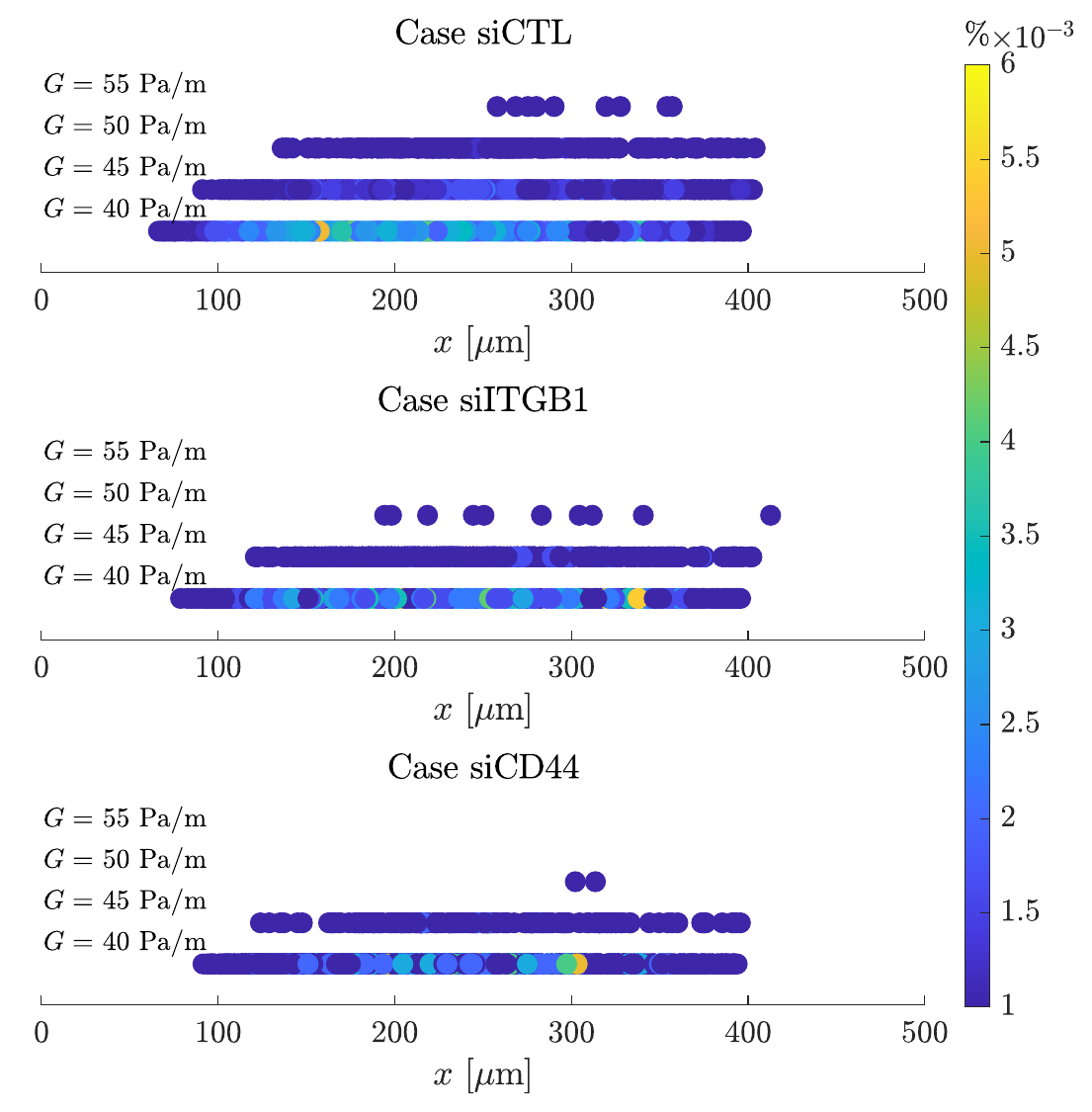}
		\caption{{Generation of $10^5$ synthetic cells considering $G= 40, 45, 50, 55 \; \SI{}{\pascal \per \meter}$ and the $3$ protein cases {of Experiment 1} with $d-r=0.62$ and $bc\sim \mathcal{N} (\mu_{bc},\sigma_{bc})$, with $\mu_{bc}$ and $\sigma_{bc}$ are taken from Table~\ref{tab:meanpop}. For each protein modification and fluid pressure gradient $G$, we represent the position at which cells velocity goes below $\SI{35}{\mu\meter\per \second}$. The percentage of cell is given in color.}}
		\label{fig:stopcells}
	\end{center}
\end{figure}

We use the mathematical model calibrated on Experiment 1 to generate synthetic data of cell velocities.
Figure~\ref{fig:cellspredict} shows as an example the time evolution of $2000$ cell velocities when $G = G^{(1)} = \SI{50.33}{\pascal \per \meter}$ for the 3 protein cases. Using the estimated values of the parameters summarized in Table~\ref{tab:meanpop}, we fix $d-r$ to the mean $0.62$ and randomly select $bc$ with given mean and standard deviation for each cohort. The opaque black lines correspond to the velocities of the $2000$ cells and the red line to their mean. Figure~\ref{fig:stopcells} depicts the positions where the synthetic cell velocities fall below $\SI{35}{\micro\meter\per \second}$ for several values of the fluid pressure gradient $G$. The color quantifies the percentage of cells in this state.

\section{Discussion} \label{sec:discussion}

\subsection{Data extraction and statistical analysis}

One difficulty in detecting cells was to select the correct {ones}. Indeed, among the cells with an apparently free trajectory, we had to pick out the CTCs without collisions, without arrests, or without problems in tracking, e.g., due to proximity to other cells. This selection obviously has an impact on the results. This selection could be automated by a more efficient tracker, especially a fully automatic tracker even for large velocities.  In addition, both the tracking method and the experimental setup only resulted in the measurement of translational cell velocities, which did not provide any information about possible rolling of the cells.

Concerning the statistical analysis of velocity data, significant differences between mean velocities in almost all of the cohorts and subcohorts (\textit{p-values} smaller than {$4\times 10^{-3}$}, see Table~\ref{table:pvalues}) illustrate the importance of the fluid velocity and of the {CD44/ITGB3 and ITGB1/ITGA5} proteins in the adhesion phenomenon. The mean velocities were not significantly different between  {siITGB1$^{(3)}$ and siCD44$^{(3)}$ ($p = 0.6$),} probably due to the higher fluid velocity impeding adhesion, and siITGB3$^{(1)}$ and siITGB3$^{(2)}$ ($p = 0.2$). This is also supported by the linear regressions on the velocities, see Table~\ref{table:linear_regression}. In each cohort and subcohort, the intercept smaller than $1$ indicates the presence of adhesion, higher for lower fluid velocities and for siCTL cases. The slopes show a decelerating dynamics for {$13$ out of $17$ subcohorts (\textit{p-values} less than $5\times 10^{-2}$)}. These values are mostly negative but close to zero, indicating the presence of stationary velocity profiles, see Figure~\ref{fig:fit}. Non-significant \textit{p-values} slopes in Experiment $2$ and $3$ are related to a lower distance $h_f^m$ of CTCs from the wall. Indeed, in this case cells have already  experienced adhesion as it comes out from the lower intercept estimate values.
The goal of the mathematical model --~presented in Subsections~\ref{subsec:fluid_velocity} and~\ref{subsec:cells_velocity}~-- was to understand and quantify these initial observations.

\subsection{Fluid modeling} \label{subsec:fluidmodeling}

As for the fluid modeling, the choice of the pressure gradient under the cosinusoidal form has an important impact, since it can be shown that in our context the imaginary part $i_f$ is larger than the real part $r_f$, which means that most of the oscillations of the cell velocities are under the cosinusoidal form. To make more complex assumptions, a better knowledge of the pump is needed.

As for the estimation of fluid parameters done mostly in the first step of the estimation process (see Subsection~\ref{ref:estimparam1st}), the strong correlation between fluid and adhesion parameters leads to identification problems. For example, we observed that multiple values of the parameter pair $(bc,\bar{u}_f(h^m_f))$ resulted in similar fits. We then selected very carefully the priors of the fluid parameters using the literature.  The prior value of $h_f^m$ (only parameter appearing in the mean velocity values $\bar{u}_f$) has a strong impact on the results. The value has been selected to be coherent with the velocity measures with single-particles performed in~\cite{follain} (see Figure~\ref{fig:mean}). Its estimation led to mean velocity values very close to the estimated ones in~\cite{osmani2}, which are $\SIlist{100;200;400}{\micro\meter\per\second}$. The prior of the angular velocity $\omega_f$ and its dependence on the pressure gradient can be derived directly from Figure~\ref{fig:vel_mean}-Top-Left (10 oscillations \SIlist{5}{\second} gives $\sim 10/5\times 2\pi \sim$\SIlist{12.6}{\per\second}). As for the correction amplitude $\xi_f$, which appears in the amplitudes $r_f$ and $i_f$ in Equation~\eqref{eq:compact_fluid_eq}, we have halved its value in each successive cohort. This is confirmed by the amplitudes observed in Figures~\ref{fig:vel_mean} and~\ref{fig:fit}. This last hypothesis was also confirmed by looking at the AIC for the case where $\xi_f$ was constant (AIC=130019.6) as opposed to $\xi_f$ depending on fluid pressure (AIC=129908.7). Of course, more complex hypotheses could be tested, but this would also require a better understanding of how the pump works.

\subsection{Adhesion modeling} \label{subsec:cellmodeling}

\paragraph{Modeling}

The adhesion dynamics is described by an ODE on the adhesion density. The modeling choice is built on Assumptions~\eqref{assump:hp1}-\eqref{assump:hp2}. Assumption~\eqref{assump:hp1} states constant binding rates in each velocity cohort. By doing so, the effects of velocity oscillations on the adhesion dynamics are neglected. Several biophysical studies investigated the relation between the load applied on a cell and its binding dynamics, and introduced catch bonds or slip bonds (see e.g~\cite{chang_state_2000,caputo_adhesive_2007,beste_pnas_2008}). 
\tr{
At a lower scale, local binding rates can be given as a function of the membrane distance to the wall \cite{zhu_cell_2000}, which could be averaged in a macroscopic rate decreasing with cell velocity. Catch bonds are well described in particular for integrins-based adhesions with a load-based reinforcement \cite{strohmeyer_fibronectin-bound_2017,bharadwaj_v-class_2017}. In our model, this can be described by either a reproduction rate $r$ or a force parameter $b$ increasing with $u_f$. Finally, increased dissociation rates with shear force (slip bonds) are commonly considered and validated in other experimental settings \cite{zhu_cell_2000, zhu_kinetics_2000, erdmann_adhesion_2004}.
However, these characterizations arise from lower scale experiments, depicting the effect of continuous variations in shear forces \cite{bongrand_ligand-receptor_1999,weisel_proteinprotein_2003,paul_calculating_2022}. In our context, it relates to the effects of fluid velocity oscillations on the adhesion dynamics. Some studies have highlighted the importance of oscillatory forces on the adhesion dynamics \cite{kong_stability_2008,irons_effect_2018,irons_switching_2020}, so that it could have been relevant in our modeling approach. In the end, our choice of neglecting fluid oscillations in the binding terms followed from the experimental data. As can be seen in Figure 7, the cell velocity oscillations were very regular, even in decelerating cells and among cells. This strongly suggested that the oscillating fluid forces variations at this timescale were not enough to disrupt the macroscopic cell regime. 
}
Assumption~\eqref{assump:hp1} then enables to derive an explicit solution for the bonds density over time given in Equation~\eqref{eq:n}. 
It was found during the estimation of parameters that $d-r > 0$. In this case, the bonds density increases exponentially and then saturates at $c/(d-r)$., while the asymptotic cell regime $d_\%$ can not reach $1$, preventing cell arrest. This shows that our modeling is not suitable to account for arrested cells.

In our model, cell velocity is given by the difference between the fluid velocity and an adhesion term. This formulation is classically found in other modeling approaches, see e.g~\cite{korn_dynamic_2008,li,grec,etchegaray}. In a macroscopic setting, the adhesion term is proportional to the closed bonds density, and involves both geometric features and forces and torques exerted by the bonds. \tr{Note that the linearity in adhesion density intertwine adhesion dynamics and force parameters.} Rather than making this term explicit, we kept a minimal framework and compared three expressions.

The constant force model, given by $u_c(t) = u_f(t) - b N(t)$, accounts for constant binding forces at the cell scale, see~\cite{etchegaray}.
The fluid-dependent force model writes $u_c(t) = u_f(t) (1 - b N(t))$ and can be seen as a simplification of a model for elastic bonds given in~\cite{grec}. 
Finally, the cell-dependent force model writes $u_c(t) = u_f(t) - b u_c(t) N(t)$, leading to Equation \eqref{eq:u_c}. This framework can be seen as a macroscopic viewpoint for the average force exerted by an elastic bond over its lifetime. Several studies show matching microscopic viewpoints involving structured bonds density capturing bonds elongation. In~\cite{li}, the membrane at the cell rear moves away from the wall at a normal velocity proportional to $u_c$. In~\cite{grec,milisic_asymptotic_2011,milisic_structured_2015}, elongation is the product of the bond's age and the cell velocity. Furthermore, in~\cite{milisic_asymptotic_2011,milisic_structured_2015}, a scaling limit in fast bonds turnover and rigid forces allowed to justify rigorously a macroscopic adhesion term proportional to $u_c(t)$. 

Assumption~\eqref{assump:hp2} consists in the choice of the cell-dependent force model over the others. \tr{Indeed, the associated deceleration term rewrites $\frac{bu_f }{1+bN}N$. From the biological viewpoint, it depicts an adhesion force proportional to $u_f$ (as for catch bonds) that is shared among bonds, as in adhesion clusters \cite{zhu_cell_2000,changede_integrin_2017}. In contrast, the two other models, less realistic, were based on additive individual adhesion forces, either constant or proportional to $u_f$. Finally, this assumption was validated by a better adequation with experimental data in the first estimation step.} The fluid-dependent force model (AIC=129910.6) slightly differs from the cell-dependent one (AIC=129908.7) in AIC, \tr{suggesting a weak importance of load sharing in the observed regime}. Using the constant force model strongly degrades the AIC (AIC=130066.8) showing that this assumption should be excluded. 

\paragraph{Parameters estimation}
Concerning the adhesion parameters, we could only estimate $bc$ and $d-r$ {for the chosen cell model}. In contrast with the fluid whose behaviour is the same in all experiments, cell parameters vary from cell to cell. They were taken as the sum of a fixed population effect and of an individual random term. Several attempts have been realised to obtain the optimal parameters from which we could interpret the biological phenomenon at study. In particular, we have considered a covariate model for $bc$ in order to integrate the cohorts effects. 
One could also add covariates with respect to subcohorts, but this seems only to complexify the definition of $bc$ without improving the AIC.
The same procedure can be applied to $d-r$. Again we have tried this strategy in the first estimation step adding covariates respect to cohorts, subcohorts and adhesion group and both fixing $d-r$ in the population and considering individual variability. We got either similar or higher AICs (superior to $129920$) in estimated parameters. Consequently, we considered only a fixed effect on $d-r$. These choices aim to reduce the number of estimated parameters, while making sure to have enough decelerating cells in each estimation group. 

\tr{Calibrating mathematical models on noisy and partial data is a challenging task. Both the model and the estimation strategy should be carefully adapted to the nature and scale of the data in order to provide a reliable framework for biological interpretation. Several of our choices aimed at ensuring the relevance of the obtained parameters. First, the preliminary estimation of fluid parameters allowed to reconstruct the better understood experimental system before investigating the cell behaviors. Moreover, the mixed-effect modeling, while increasing the number of parameters, constrains them with the knowledge that the observed cells behave following a common framework. This approach improves identifiability and reduces the risk of obtaining spurious parameter sets in the population. Some results were also indicative of coherent estimations: the obtained cell heights were no more than $\SI {1.2}{\micro\meter}$ apart one from another, and they were in accordance with the expected values from the experimental settings. Furthermore, while the three experiments were conducted independently, the estimation results allowed to recover the known molecular associations across them, at least fully for the lower fluid velocity. Altogether, these estimations provide a credible framework with respect to the biological phenomenon under study.}

\paragraph{Interpretation of the estimated adhesion parameters}
A better adhesion parameters interpretability is given by the percentage of decrease in cells velocity $d\%$, depending both on $bc$ and $d-r$. Figure~\ref{fig:100200400modif} shows significant differences for low velocities in all the three experiments. As fluid velocity increases, we loose significant differences since adhesion is less important. This also justifies the interest in reproducing experiments under lower fluid velocities through the digital twin as in Figure~\ref{fig:stopcells}.

In Figure~\ref{fig:MODIFvelocity}, a comparison is done over different fluid velocities. In particular we compare control cases and similar proteins over the three experiments. Significant differences are observed whenever adhesion is meaningful, \textit{i.e.} control (red boxes) and stronger adhesion protein depletion (green boxes). This stands for a more important contribution in the adhesion phenomenon of the first -- albeit weak-- protein CD44 as well as ITGB3.

Moreover, in both figures we can also compare the different control cases effect (which is similar as expected) as well as proteins one. Proteins CD44 and ITGA5 (siITGB1 and siITGA5 case respectively) have an active role in the first steps of adhesion, whereas ITGB1 and ITGB3 (siCD44 and siITGB3 case respectively) has a minor impact.

As stressed before, this analysis has been pursued for $d-r$ fixed. In this context, individual variability does not change the AIC, instead it can cause the loss of significant differences among proteins adhesion effects. \tr{A constant value indicates that the timescale of deceleration (or equivalently of global adhesion growth) is identical in each case.} In conclusion, the choice on $d-r$ estimation is based on better decelerating results in the second estimation step. Note that deciphering further experiments with a wider fluid velocities range or a longer observation duration could require relaxing this choice. \tr{Consequently, the cell variability between cohorts and subcohorts is carried by the binding term $bc$.}

\subsection{Comparison with the literature}

\tr{
Conclusions made from the estimation results are tightly linked to the experimental setting, considered cells and proteins, and scale of measures (see e.g \cite{giavazzi_rolling_1993} for variable behaviors among tumor cells in flow). One can nonetheless discuss their coherence with respect to the literature. 
A deceleration timescale independent of shear rate was also found in \cite{cheung_adhesion_2011} for other adhesion bonds. Note that this is not in contradiction with the concepts of catch and slip bonds, since having an increasing dissociation rate $d$ and a decreasing reinforcement rate $r$ with respect to the fluid velocity may maintain a constant difference $d-r$ if the magnitudes are similar. However, the scaling of present experiments were not suited for capturing these features. }

\tr{
In our work, we find that the fast and metastable CD44/ITGB3 bonds are more important for CTCs deceleration \emph{in vitro} than slower and more stable ITGB1/ITGA5 ones, which could not be established in the prior analysis. Nevertheless, the \emph{in vivo} experiments on the zebrafish embryo had underlined that only the first type of bonds was able to induce early cell arrest, indicating a higher adhesion efficacy at faster flows \cite{osmani}. The importance of CD44 in adhesion is already established in more general settings \cite{honn_adhesion_1992,degrendele_cd44_1996,katayama_cd44_2005,thomas_carcinoembryonic_2008,mcever_rolling_2010}.
We further find that the binding parameter $bc$ is a decreasing function of fluid velocity in the control case, and weaker in the siITGB1 case, but not in the siCD44 case. As a result it can be considered as a feature of CD44-based adhesion, yielding increasing force parameter or binding rate with decreasing velocities. Note that opposite trends have also been observed in other context, such as an increase in CD44-based adhesion of leukemic cells above a shear stress threshold \cite{christophis_shear_2011}. It is also known that CD44 has a larger binding rate than ITGB1, but forms weaker and shorter-lived bonds \cite{osmani}. In combination with our findings, a possible framework could be that, at a given fluid velocity level, CD44 is associated with larger dissociation and reinforcement rates than ITGB1, and that the larger binding rate compensate for the weaker forces. Conversely, at the same fluid velocity level, ITGB1/ITGA5 bonds may be associated with larger forces but much lower binding rates, together with lower dissociation and reinforcement rates yielding close values for $d-r$. Integrins being tightly linked with cell cytoskeleton, they are particularly responsive to load and involved in many biological processes involving mechanotransduction \cite{puklin-faucher_mechanical_2009,roca-cusachs_finding_2012,huaman2020molecular}. They are primarily considered as mediating firm adhesion, and behave as catch-slip bonds within the relevant speed range \cite{mcever_rolling_2010,rakshit_biomechanics_2014,strohmeyer_fibronectin-bound_2017,bharadwaj_v-class_2017}. Altogether, our findings are therefore compatible with low scale studies on cell adhesion.
}

\tr{
	Inflow cell adhesion has been investigated in various \emph{in vitro} settings for fibroblasts or immune cells \cite{truskey_effect_1990,yamamoto_quantitative_2000,dong_biomechanics_2000,simon_leukocyte_2002}. Few microfluidics studies have investigated tumor cells adhesion. In the absence of flow, cell types heterogeneity in adhesion strenght has been pointed out in single cell measures in \cite{mao_adhesion_2018}. 
	 The role of flow rate and acceleration on resting tumor cells detachment in a microchannel has been highlighted in \cite{cheung_detachment_2009}. The importance of chemokines in extravasation has been studied in \cite{song_microfluidic_2009,riahi_microfluidic_2014}. Finally, two microfluidic approaches investigated the adhesion dynamics of MDA-MB-231 breast cancer cells \cite{cheung_adhesion_2011,rupprecht_tapered_2012}. In \cite{rupprecht_tapered_2012}, a wide range of shear stresses applied on $500$ to $1500$ resting cells showed that the shear stress threshold for detachment was unchanged among all conditions, whereas detachment kinetics were different. In \cite{cheung_adhesion_2011}, cells were flowing and interacting with an anti-EpCAM ligands-coated wall.}
Sequential fitting of the computational model of~\cite{korn_dynamic_2008} on mean translational velocities for several shear rates allowed to identify the average cell height, binding force, and bonds spring constant. Normalized cell velocities for all shear rates were successfully fitted by a generic exponentially decreasing curve, suggesting a strong dependence of the cell velocity magnitude on the fluid velocity, whereas it was not the case for the typical decay time. Several differences exist between our frameworks. From the biological viewpoint, the wall in~\cite{cheung_adhesion_2011} is passive, while ours is a monolayer of endothelial cells, whose flow-driven active behaviour in cell arrest has been observed~\cite{osmani}. Together with lower fluid velocities, it may explain their measures of smooth velocity decays until cell arrests, that were not observed in our case. {Moreover,  the EpCAM-based approach induces a bias towards cells with strong epithelial identity, which are not necessarily the most aggressive (see e.g~\cite{hyenne_liquid_2022,peralta}).} Furthermore, we worked with partial observations since cell velocities were not measured from their entrance in the experimental setting. On the other hand, we {took into account} the {temporally} oscillating fluid velocity that {needed} to be reconstructed and developed a mixed-effects calibration strategy that was able to deal with the individual cell velocities over time {and to estimate the parameters jointly}. In this setting, we obtained insights on the role of the fluid velocity that are consistent with previous observations~\cite{cheung_adhesion_2011,follain,osmani}.  Finally, our original and robust approach allowed not only to investigate the respective roles of several adhesion proteins in the cell dynamics, but also to build a digital twin of flowing cells with adhesion in other fluid configurations.

Therefore, our methodology represents a valuable contribution not only to validate experimental results, but also to interpret them more finely and use them to build further knowledge.

\section{Conclusion and perspectives} \label{sec:conclusions}

In this work, we have attempted to characterise CTCs in the flow and their interaction with the vessel wall, relying on the \textit{in vitro} experiments performed by Osmani and collaborators in~\cite{osmani,osmani2}. Whereas previous analyses focused on cell arrest, the use of the CSRT tracker allowed us to record trajectories and velocities of individual cells.
We were able to analyse {$3$ experiments comprising} different cell cohorts with respect to three different values of fluid pressure gradient (below the threshold for efficient CTC adhesion found by Osmani and collaborators in~\cite{follain}) and $7$ different protein expressions (siCTL, the control case relative to each of the three experiment; siITGB1, depletion of ITGB1, integrin that promotes adhesion stabilisation; siCD44, depletion of CD44, protein involved in early adhesion; siITGA5, depletion of ITGA5, whose effect is similar to ITGB1; siITGB3, depletion of ITGB3, with a comparable role to CD44). Statistical analysis of the mean of the extracted cell velocities and linear regression allowed the observation of a slowing behaviour over time, see Tables~\ref{table:pvalues} and~\ref{table:linear_regression}. This shows that adhesion is a continuous-time phenomenon involving CTCs in a fluid with a velocity below the threshold of $\SI {400}{\micro\meter\per\second}$.

Since the fluid velocity was not measured directly, we only knew the values of the pressure gradient generated by the peristaltic pump that made up the device. This lack of data was compounded by our lack of knowledge about the pump. However, we were able to establish a Poiseuille regime and describe the fluid velocity as a combination of oscillatory functions induced by the pump and evident in the tracked cell velocity in Figure~\ref{fig:vel_mean}.
We then focus on the modeling of the cell velocity. The oscillating Poiseuille flow was weakly coupled to a simple ODE model for cell adhesion that describes the cell velocity as the fluid velocity affected by bond formation and disruption.

Optimal parameters for our model were not easy to find. Indeed, there are practical problems with identifiability, mainly due to data noise and little information about the fluid parameters. Our strategy to overcome this problem is based on a mixed-effects model and careful selection of fluid parameter priors.

The well-designed parameter estimation has led to very attractive results. The first important result is that the estimated parameter values allow deciphering the CTC binding and deriving biological outcomes. We briefly summarize them here: (1) the strong influence of fluid velocity on protein binding: a low fluid velocity favours a decrease in cell velocity and the formation of bonds, while a high fluid velocity makes it difficult to observe this adhesion phenomenon even when all adhesion proteins are expressed, (2) the importance of CD44/ITGB3 in early adhesion compared to the integrin ITGB1/ITGA5.
Similar conclusions have already been assessed for  point~(1) using statistical strategies. However,  point~(2) was only assessed using \textit{in vivo} experiments and could not be extracted from the \textit{in vitro} experiments prior to our work. This underlines the quality of the strategy we developed, which is based on mathematical modeling and data assimilation.

This methodology leads to a second important result. Indeed, since it is based on a mathematical model --~once the model parameters have been estimated~-- the digital twin of the setup can be reproduced, making it possible to learn more details about the behavior of the cells, even for configurations that were not considered during the experiments.

{In conclusion, it can be noted that this experimental setup is quite simple and far from reality, but from a mathematical point of view, its adoption allows the selection of a model and its parameterization. It should be noted that even in this case, where a lot of data can be extracted, some parameters are not identifiable. With less data, such as only the percentage of locally arrested cells, as is often the case with \emph{in vivo} data, it would be impossible to identify the estimated parameters. This first step is therefore crucial. Indeed, these first estimates make it possible to establish laws for the parameters as a function of fluid velocity and to perform simulations even for configurations not considered in the experiments, thus paving the way for \emph{in vivo} data.}

As for the perspectives, the first one concerns the improvement of the CTC tracker from the experimental videos, since it is not fully automatic and has a great need of optimization. The second is the development of a mathematical model adapted to the \textit{in vivo} experiments. These data should allow us to incorporate cell arrest into our model. The more complex geometry will require a more complex model of blood circulation. 
 {If we consider a different experimental setup, this model of flow velocity should be adapted based on the very extensive literature on the subject. In particular, the fact that blood is a non-Newtonian fluid --~as its viscosity depends on the shear rate~-- must be taken into account}.
Finally, from a biological standpoint, it is possible to use the model presented to study and predict additional molecular modes involved in the arrest of CTCs at the vascular wall.

\paragraph*{Author contributions}
AC and CE designed the study based on the biological data generated by NO in the team of JGG. GC and JG analyzed the data. AC and GC implemented the software code. AC, CE and GC interpreted the results. AC, CE and GC wrote the manuscript and JG the supplementary materials.

\section*{Acknowledgements}
The work of JGG and NO has been funded by Plan Cancer 2014-2019 (OptoMetaTrap), CNRS IMAG’IN (JGG) and by institutional funds from INSERM and the University of Strasbourg.

\section*{Declaration of interest}
The authors declare no competing interests.

\bibliographystyle{unsrtnat}
\bibliography{references}

\newpage
\appendix

\title{Supplementary Materials of <<\commandtitle>>}
\setcounter{equation}{0}
\setcounter{figure}{0}
\setcounter{table}{0}
\setcounter{page}{1}
\makeatletter
\maketitle

\section{Numerical modeling of the fluid}
\label{appendix:fluid}

\begin{figure}
	\begin{center}	
	\hspace*{-0.1cm}
		\includegraphics[width=\textwidth]{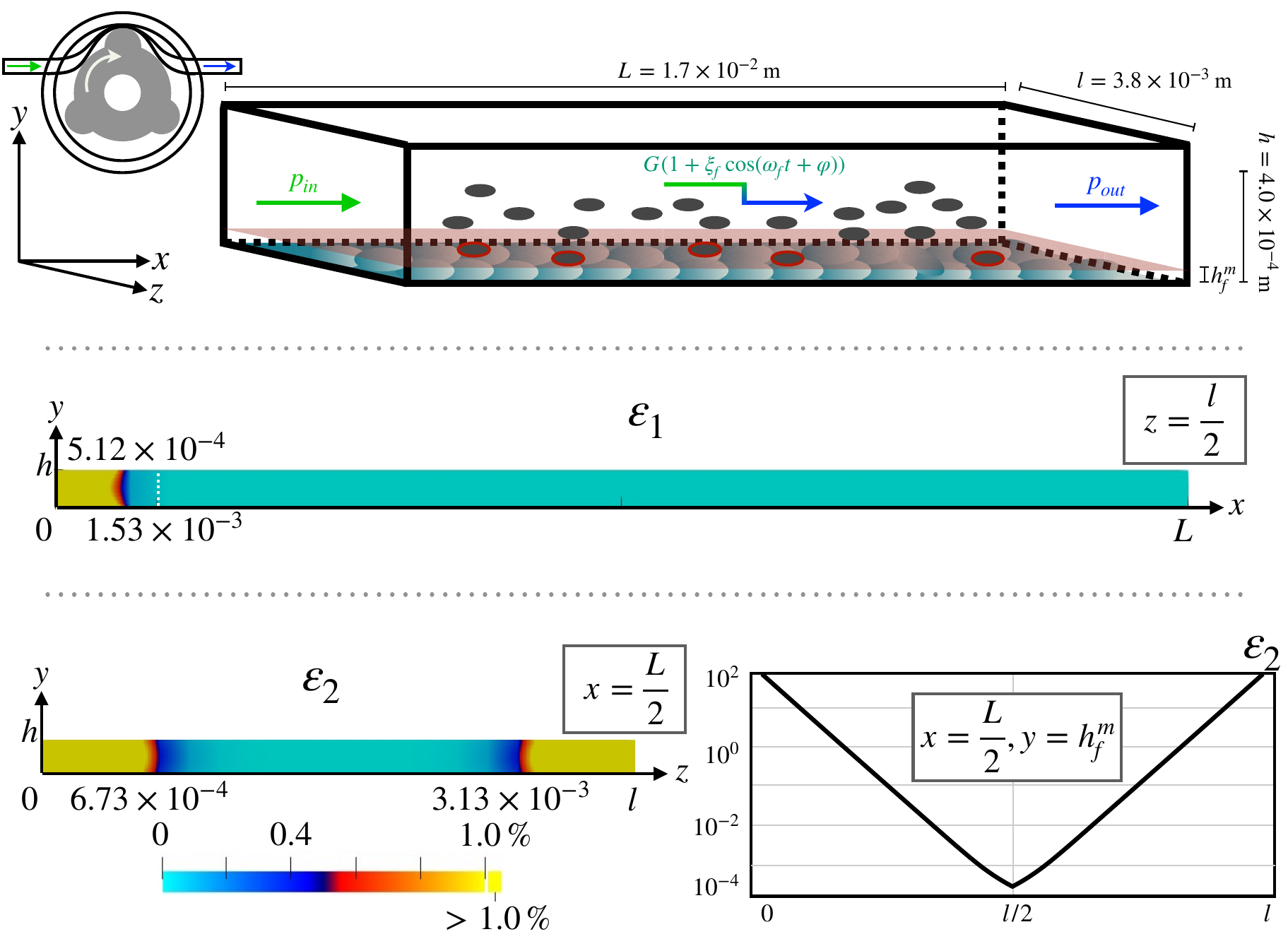}
		\caption{Top: Experimental setup. Middle: Relative error $\varepsilon_1$ on the cross-section  $(0, L) \times (0, h) \times \left\{ \frac{l}{2} \right\} $. For $x\geq  \SI{5.12d-4}{\meter}$ we have, $\varepsilon_1(\mathbf{\cdot}) \leq 1 \%$. For reference, the entry length $\ell =\SI{1.53d-3}{\meter}$ given by~\cite{Ferreira2021} is also displayed. Bottom-Left: Relative error $\varepsilon_2$ on the cross-section $\{\frac{L}{2}\} \times (0,h) \times (0,l)$. 
		For $z\in [ \SI{6.73d-4}{\meter}, \SI{3.13d-3}{\meter} ]$ we have, $\varepsilon_2(\mathbf{\cdot}) \leq 1 \%$. Bottom-Right: Relative error $\varepsilon_2$ on the cross-section $\{\frac{L}{2}\} \times \{h_f^m\} \times (0,l)$ }
		\label{fig:freefem1}
	\end{center}
\end{figure} 

\subsection{Validation of the hydrodynamic entrance length $\ell$}
\label{subappendix:hydrolength}
Let us define the domain $\Omega := ( 0, L ) \times (0, h ) \times ( 0, l ) $,  $\partial{\Omega}$ its boundary and $\partial{\Omega}_{in} := \{0 \} \times (0,h) \times (0,l) $ the boundary corresponding to the channel inlet. We would like to numerically establish the distance $\ell$ (along the $x-$axis) at which a uniform velocity profile from $\partial{\Omega}_{in}$ fully develops into a Poiseuille profile. To do so, we compare the solution of the two following problems: (i)~a Poiseuille profile is already developed over the entire domain $\Omega$, (ii)~boundary layer equations are used to take into account the entrance effect near the inlet. Let $u_P$ be the fluid velocity anywhere in $\Omega$ according to the Poiseuille equations 
\begin{spreadlines}{1ex}
\begin{equation}
\begin{dcases}
         \rho \frac{\partial u_P}{\partial t} - \mu \Delta u_P  = G (1 + \xi_f \cos(\omega_f t + \varphi))   \quad &t>0, \; \Omega, \\
         u_P(t,\cdot) = 0  \quad &t>0, \; \partial{\Omega}, \\ 
         \label{eq:annexpoiseuille}
         u_P(0,\cdot) = 0  \quad & \Omega,
\end{dcases}
\end{equation}
\end{spreadlines}
and $u_B$ be the fluid velocity anywhere in $\Omega$ according to the boundary layer equations 
\begin{spreadlines}{1ex}
\begin{equation}
\begin{dcases}
         \rho \left(\frac{\partial u_B}{\partial t} + u_B \nabla \cdot u_B \right) - \mu \Delta u_B  = G (1 + \xi_f \cos(\omega_f t + \varphi))   \quad &t>0, \; \Omega, \\
         u_B(t,\cdot) = 0  \quad &t>0,\; \partial{\Omega} \backslash \partial{\Omega}_{in}, \\
         \label{eq:annexboundary}
         u_B(t,\cdot) = \max_{t>0} u_P \quad &t>0,\; \partial{\Omega}_{in}, \\
         u_B(0,\cdot) = 0  \quad & \Omega.
\end{dcases}
\end{equation}
\end{spreadlines}
The third equation of Problem~\eqref{eq:annexboundary} corresponds to a uniform velocity profile at the inlet of the channel, which is here equal to the maximum velocity of the fluid with a fully developed Poiseuille profile. This choice was made to be consistent with the steady state hypothesis.
Problems \ref{eq:annexpoiseuille} and \ref{eq:annexboundary} are numerically solved using \texttt{FreeFEM++}~\cite{freefem}. The domain is discretized on a $120 \times 10 \times 20$ mesh, using P1 Lagrange elements. The timestep $\Delta t$ is taken to be $\SI{0.5}{\second}$ and the total time of the simulation is $ \SI{15}{\second}$.

For the parameters, we work with the \textit{worst case scenario}, \textit{e.g.} the highest Reynolds number possible, with $Re = 3.75 $. We recall that $L = \SI{1.70d-2}{\meter}$, $h =\SI{4.00d-4}{\meter}$, $l = \SI{3.80d-3}{\meter}$, $\rho = \SI{1.00d3}{\kilo \gram \per \cubic \meter}$, $\mu = \SI{7.20d-4}{\pascal \second}$, $G = \SI{201.32}{\pascal \per \meter}$, $\omega_f = \SI{4.88d1}{\radian \per \second}$, and $\varphi = \SI{0}{\radian}$, see Subsections \ref{subsec:data_availability},~\ref{subsec:fluid_velocity} and~\ref{subsec:param_estimation_results} for details.
By convention, a flow can be considered fully developed when its velocity profile matches the asymptotic one with an error margin of less than a percent.
Assuming $u_B(t,\cdot) \ne 0$, for all $t >0$, the relative error in percent is  computed as follows 
\begin{align*}
    \varepsilon_1 (\cdot):= \max_{t>0} \left ( \dfrac{|u_B(t,\cdot) - u_P(t,\cdot) | }{|u_B(t,\cdot) | } \right) \times 100, \quad \Omega.
\end{align*}

Figure~\ref{fig:freefem1}-Middle shows the relative error $\varepsilon_1$ over the longitudinal cross-section $(0, L) \times (0, h) \times \left\{ \frac{l}{2} \right\} $. 
One may remark the hydrodynamic entrance length $\ell$ --~displayed along the cross-section~-- seemingly over-predicts the distance at which $\varepsilon_1$ drops below one percent. The numerical entrance length at which $\varepsilon_1 \leq 1\%$ is given by $\ell_{num}=\SI{5.12d-4}{\meter}$. Numerical tests including spatial and temporal convergence were  performed to validate the value of $\ell_{num}$.


From Subsection~\ref{subsec:fluid_velocity}, we know the data was obtained $\frac{L}{2}=\SI{8.50d-3}{\meter}$ away from the inlet, which is a full order of magnitude above the upper bound $\ell$ of the hydrodynamic entrance length. Under those circumstances, the velocity profile of $u_B$ closely matches the velocity profile of $u_P$ and is considered as a fully developed Poiseuille flow.

\begin{remark}
The value $\ell = \SI{1.53d-3}{\meter}$ given in Subsection~\ref{subsec:fluidmodeling} and coming from~\cite{Ferreira2021} is superior to our numerical evaluation of the hydrodynamic entrance length $\ell_{num}$. Furthermore, additional hydrodynamic entrance length comparisons were performed with other formulas available in the literature (see~\cite{Ferreira2021}, Table 1), and all of them were found to overpredict $\ell_{num}$. This difference is explained by the relation between the velocity of the fluid far from the inlet and its velocity at the inlet boundary. As shown in~\cite{Ferreira2021}, their study was performed for a ratio whose maximum is between $1.5$ and $2$, while in our study we set this maximum at $1$. Therefore, their entry velocity is much higher than the asymptotic velocity of the fluid, which is reflected in the reported value of $\ell$.

\end{remark}

\subsection{Validation of the fluid expression }
\label{subappendix:1Dfluid}

In Subsection~\ref{subappendix:hydrolength}, we have shown the device was \textit{long enough} for a Poiseuille flow to become fully developed and therefore independent on the $x-$axis. While this reduces Problem~\eqref{eq:annexpoiseuille} into a 2D one, we can simplify it even further by showing the channel is \textit{wide enough} that we can neglect a change in the $z-$axis as well. We recall that aspect ratio $AR = \frac{h}{l}$ equals to $\SI{1.05d-1}{}$, which tells us the microfluidic device is roughly $10$ times wider than it is tall. 
To show that 1D reduced model (in $y$-axis direction) is reasonable, we compare the fluid velocity $u_P$ to the expression $u_f$, which must verify 
\begin{spreadlines}{1ex}
\begin{equation}
\begin{dcases}
         \rho \frac{\partial u_f}{\partial t} - \mu \frac{\partial^2 u_f}{\partial y^2}  = G (1 + \xi_f \cos(\omega_f t + \varphi))   \quad &t>0, \; (0,h), \\
         u_f(t,\cdot) = 0  \quad &t>0, \; (0,h), \\
         \label{eq:annexpoiseuille1D}
         u_f(0,\cdot) = 0  \quad & (0,h).
\end{dcases}
\end{equation}
\end{spreadlines}
Using the same parameter values and mesh as in Subsection~\ref{subappendix:hydrolength}, we solve Problem~\eqref{eq:annexpoiseuille1D} with \texttt{FreeFEM++}. The same error threshold of one percent is taken, with 
\begin{align*}
    \varepsilon_2 (\cdot):= \max_{t>0} \left ( \dfrac{|u_P(t,\cdot) - u_f(t,\cdot) | }{|u_P(t,\cdot) | } \right) \times 100, \quad (0,h) \times ( 0, l ).
\end{align*}

Figure~\ref{fig:freefem1}-Bottom-Left gives $\varepsilon_2$ on the lateral cross-section $ \{\frac{L}{2}\} \times (0,h) \times (0,l)$ corresponding to the section from which the data were obtained.  For $z\in [ \SI{6.46d-4}{\meter}, \SI{3.15d-3}{\meter} ]$ we have, $\varepsilon_2(\cdot) \leq 1 \%$.
Figure~\ref{fig:freefem1}-Bottom-Right shows the relative error $\varepsilon_2$ at the camera's focal plane, using the value of $h_f^m$ found in Subsection~\ref{subsec:param_estimation_results}. As expected from the observation made in Figure~\ref{fig:freefem1}-Bottom-Left, the distance between the two inner ticks does not increase nor decrease in any significant way. The sub$-1\%$ relative error area thus spans most of the microfluidic device, with $\{\frac{L}{2}\} \times (0,h) \times (\SI{6.46d-4}{},\SI{3.15d-3}{}) $. Upon closer inspection of the data provided in~\cite{osmani,osmani2}, we know the data was obtained at the middle of the device length wise, but also width wise, as the lateral sides of the device are not visible in the videos. Given the width of the sub$-1\%$ area, $u_P$ can therefore be represented by~$u_f$.

\subsection{Validation of the parabolic shape for the velocity profile}
\label{subappendix:womersley}
For a purely oscillating pressure gradient, a parabolic velocity profile can be considered for a $\mathbf{Wo}$ up to $1$, after which the profile rapidly evolves into a \textit{plug}-like shape~\cite{womersley1955method}. Counting the number of oscillations in Figure~\ref{fig:vel_mean}, it can be seen that $\omega_f$ depends on the cohort. The highest value is reached at the third cohort and is close to \SIlist{50}{\per\second} ($\sim$12 oscillations on \SIlist{1.5}{\second} gives $\sim 12/1.5\times 2\pi \sim$\SIlist{50}{\per\second}), resulting in a $\mathbf{Wo}$ around $1.6$. However, it is shown in~\cite{ma} that a parabolic profile can still emerge for moderately low values of $\mathbf{Wo}$ ($\approx 2$) when
\begin{align*}
    |\text{Re}(u_f(t,y))| \gg |\text{Im}(u_f(t,y))| \qquad t>0, \quad y\in (0,h).
\end{align*}

\begin{figure}
	\begin{center}	
		\includegraphics[width=0.55\textwidth]{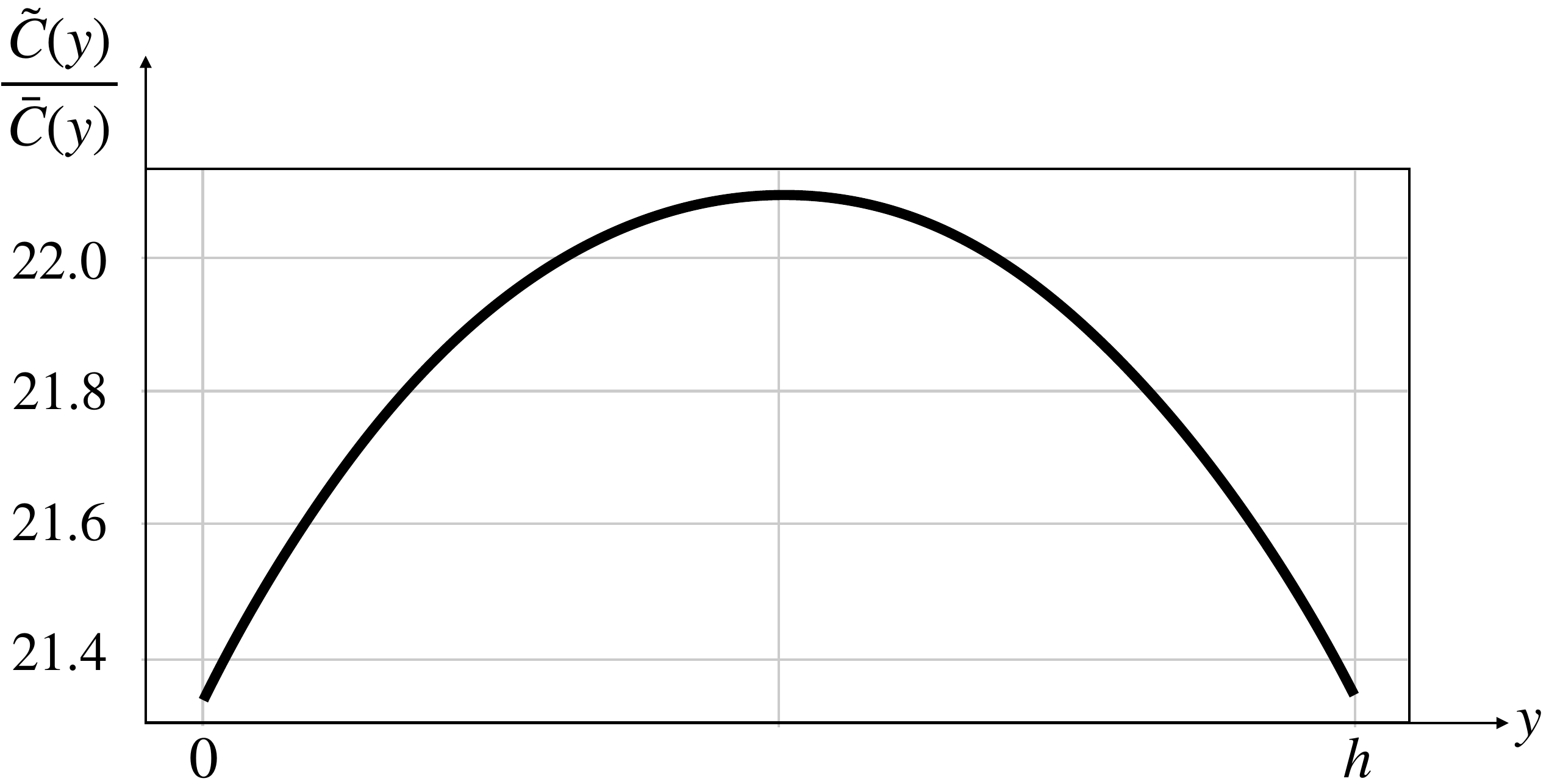}
		\caption{Quotient $\dfrac{\overline{C}(y)}{\widetilde{C}(y)}$ for $y\in (0,h)$.} 
		\label{fig:c_ctilde}
	\end{center}
\end{figure} 

Using the superposition principle, one can rewrite $u_f$ as a sum of its constant and oscillating part, that we respectively denote by $\overline{u}_f(t,y)$ and $\widetilde{u}_f(t,y)$. We now have to check that
\begin{align}
    |\text{Re}(\widetilde{u}_f(t,y))| + |\text{Re}(\overline{u}_f(t,y))| \gg |\text{Im}(\widetilde{u}_f(t,y))| + |\text{Im}(\overline{u}_f(t,y))| \qquad t>0, \quad y\in (0,h)\,. 
    \label{cond:parabolic}
\end{align}
From Equation~\eqref{eq:compact_fluid_eq}, one has $|\text{Re}(\overline{u}_f(t,y))| = \overline{C}(y) > 0$ and $|\text{Im}(\overline{u}_f(t,y))| = 0$, so that the condition rewrites
\begin{align*}
    |\text{Re}(\widetilde{u}_f(t,y))| + \overline{C}(y) \gg |\text{Im}(\widetilde{u}_f(t,y))|  \qquad t>0, \quad y\in (0,h)\,,
\end{align*}
which is true for all $t$ if $\overline{C}(y)$ is taken large enough. In the most pessimistic case, we have $|\text{Re}(\widetilde{u}_f(t,y))|=~0 $ and  $\displaystyle \max_{t>0}|\text{Im}(\widetilde{u}_f(t,y))| = \widetilde{C}(y) > 0$. Therefore, it is enough to show that 
\begin{align*}
     \dfrac{\overline{C}(y)}{\widetilde{C}(y)} &\gg 1.
\end{align*}

Figure~\ref{fig:c_ctilde} shows the value of $\frac{\overline{C}(y)}{\widetilde{C}(y)}$ for $y\in (0,h)$. It can be deduced that Condition~\eqref{cond:parabolic} is always verified. This means that for $u_f = \overline{u}_f + \widetilde{u}_f $, the constant component is large enough to ensure that the magnitude of oscillations does not disrupt the parabolic shape of the velocity profile. This therefore validates our hypothesis.

\end{document}